\newcommand{\rem}[1]{}
\documentclass{amsart}
\usepackage{amsfonts,amssymb,amsmath,amsthm,mathrsfs}
\usepackage{url}
\usepackage{graphicx}
\newtheorem{thrm}{Theorem}[section]
\newtheorem{lem}[thrm]{Lemma}
\newtheorem{prop}[thrm]{Proposition}

\newtheorem{remark}[thrm]{Remark}
\theoremstyle{definition}

\def\XXint#1#2#3{{\setbox0=\hbox{$#1{#2#3}{\int}$}
     \vcenter{\hbox{$#2#3$}}\kern-.5\wd0}}

\begin{document}
\author[C.A.Mantica and L.G.Molinari]{Carlo Alberto Mantica and Luca Guido Molinari}
\address{C.~A.~Mantica and L.~G.~Molinari (corresponding author): 
Physics Department Aldo Pontremoli,
Universit\`a degli Studi di Milano and I.N.F.N. sezione di Milano,
Via Celoria 16, 20133 Milano, Italy.\\
Orcid: 0000-0001-5638-8655 and 0000-0002-5023-787X}
\email{carlo.mantica@mi.infn.it, luca.molinari@mi.infn.it}

\title[]{
 The covariant approach to static spacetimes \\
in Einstein and extended gravity theories} 

\begin{abstract}  We present a covariant study of static space-times, as such and as solutions of gravity theories.
By expressing the relevant tensors through the velocity and the acceleration vectors that characterise static space-times, the field equations provide a natural non-redundant set of scalar equations. The same vectors suggest the form of a Faraday tensor, that is studied in itself and in (non)-linear electrodynamics. \\
In spherical symmetry, we evaluate the explicit expressions of the Ricci, the Weyl, the Cotton and the Bach tensors. Simple 
restrictions on the coefficients yield well known and new solutions in Einstein, $f(R)$, Cotton and Conformal gravity, with or without charges, in vacuo or with fluid source.\\
\end{abstract}
\date{07 august 2023}
\maketitle
\section{Introduction}
The field equations of gravitational theories are covariant, and equate a geometric tensor
(e.g. Einstein, Cotton, Bach tensor) to a tensor describing matter. Solutions are usually found in coordinates that exploit the symmetries. However, there are advantages in keeping the coordinate-free tensor description as far as possible. 
Besides the formal elegance, it naturally addresses scalar identities. 
We do so in this study of static space-times, beginning with local geometry and then discussing gravity.

A covariant characterization of a static space-time involves an equation for a time-like velocity $u_k$ with a closed space-like acceleration $\dot u_k=u^j\nabla_j u_k$. The two defining vectors are the natural start for the expansion of the relevant
tensors. In absence of symmetries they are complemented by two others. This is basically the spirit of the 1+1+2 formalism introduced by Clarkson and Barrett \cite{Clarkson03}\cite{Clarkson07}. Here the second vector is fixed by the 
context. Carloni used the formalism to specify the stress tensor, and the Ricci tensor ensuing from the Einstein equation, in spherically symmetric metrics \cite{Carloni14}. In this work the two unspecified orthogonal vectors are not explicitly required, as the Ricci tensor is
constructed via the integrability conditions and linked to the electric part of the Weyl tensor.

With the vectors $u_j$, $\dot u_j$ and an orthogonal space-like pair, we consider the antisymmetric tensor ($\eta=\dot u^k \dot u_k$)
$$ F_{jk} =\mathbb E \tfrac{1}{\sqrt\eta} (u_j \dot u_k - \dot u_j u_k ) + \mathbb B (y_jz_k-y_kz_j) $$ 
and obtain the conditions on $\mathbb E$ and $\mathbb B $ to yield a Faraday tensor. The terms correspond to the electric and magnetic fields. In the Einstein theory, the equations of linear and non-linear electrodynamics constrain the Ricci tensor to a simple structure, with coefficients $R$ and $R^\star$. The weak
energy condition imposes a non-negative spatial curvature scalar, $R^\star \ge 0$, while the scalar curvature $R$ is zero 
if and only if the electrodynamics is linear. 

After the general setting, we turn to the much studied spherically symmetric static space-times, with line element
\begin{align*}
ds^2 = - B(r) dt^2 +\frac{dr^2}{B(r)} +r^2(d\theta^2 + \sin^2\theta \, d\phi^2).  
\end{align*}
The early solutions in General Relativity, named after Schwarzschild, de Sitter, Reissner and Nordstr{\o}m, 
belong to this class. In the years many others were found as black hole (BH) or compact star solutions, that fit in the present covariant description.

Beginning with geometry, we obtain the relevant static spherical tensors (Ricci, Weyl, Cotton, Bach)
as combinations of  $u_i u_j$, $g_{ij}$ and $\dot u_i \dot u_j$. Remarkably, also the energy-momentum 
tensor of linear and non-linear electrodynamics is a combination of the same elementary tensors, with 
the magnetic term $\mathbb B$ being forced to be the field of a magnetic monopole. 
By constraining the tensor coefficients to simple forms, we obtain scalar equations that recover the early metrics and others.

By the equality inherent gravity theories mentioned in the beginning, the geometric Ricci, Cotton and Bach tensors 
fix the form of the energy-momentum tensor respectively in Einstein, Cotton and Conformal gravity.
A similar construction is made in $f(R)$ theory, where the Ricci tensor and the Hessian provide the form of the matter tensor. In these four theories, it has the form of energy-momentum tensor of an anisotropic fluid or of linear or 
nonlinear electrodynamics.\\
The identification of the 
geometric and physical coefficients of the tensors in the left and right sides of the field equations, 
provides scalar equations. 
The functions $B(r)$ found on geometric grounds specify as solutions of field equations.

In a century of gravity theories, many static spherical solutions were found. This is a very short and partial recount. \\
In 1968 James Bardeen (son of John B. of BCS theory) obtained the first singularity-free BH solution of the Einstein equation \cite{Bardeen68}\cite{Ansoldi}: 
$$ B(r) = 1- \frac{2M r^2}{(r^2 + g^2)^{3/2}} $$
Ayon-Beato and Garcia reinterpreted it as a magnetic monopole solution in Einstein-non-linear electrodynamics \cite{Ayon00}.\\ 
In 2003 Kiselev published a new exact solution of the Einstein equation for quintessential matter surrounding a BH \cite{Kiselev03}
$$ B(r) = 1 -\frac{2M}{r} -\frac{K}{r^{1+3w}} $$
It raised a debate, until Visser showed in 2020 that the Kiselev BH is neither a perfect fluid nor a quintessence \cite{Visser20}. Generalisations of Kiselev space-times were recently used in the framework of gravitational lensing \cite{PAP23}. \\
In 1996 Hayward \cite{Hayward96} discovered a line element describing the local formation of a BH out of vacuum, its Bardeen-like static quiescence and final evaporation.

Bronnikov \cite{Bronnikov01} showed that Einstein gravity coupled to nonlinear electrodynamics has nontrivial spherical solutions with global regular metric if and only if the electric charge is zero and the Lagrangian 
$\mathscr L(F)$ has a finite limit as $F\to \infty$. 
In the same context, Dymnikowa \cite{Dymnikova04} studied the 
existence of regular spherically symmetric electrically charged solutions. The effects of torsion were considered by
Cotton \cite{Cotton21}.

Gravastars (gravitational vacuum stars) were introduced in 2001 \cite{Mazur23} as an alternative to BH that avoid the problems associated with horizons and singularities. Models in nonlinear electrodynamics were constructed by Lobo and Arellano \cite{Lobo07}. 

Among a great variety of spherical metrics in Einstein gravity, we quote the Yukawa BH \cite{Mazharimousavi19}, the Van der Waals BH \cite{Rajagopal14}, the global monopole \cite{Barriola89}, the Rindler--Grumiller metric \cite{Grumiller}, the logotropic BH
\cite{Capozziello23}, quantum corrections to Reissner-Nordstr{\o}m BH \cite{Wu22}. 
The study of non-linear electrodynamics coupled to $f(R)$ gravity was started by Hollenstein and Lobo \cite{Hollenstein08} \cite{Rodrigues16}.

In 1989 Mannheim and Kazanas \cite{MannKaz} obtained an exact vacuum solution of Conformal gravity, and applied it to describe the rotation curve of galaxies without dark matter
$$ B(r) = -\frac{1}{r}\beta (2-3\beta\gamma) + (1-3\beta\gamma) + \gamma r - \kappa r^2 $$
Topological black holes in conformal gravity were studied by Klemm \cite{Klemm98}.

Based on the conformal action, but with variation in the connection, a theory named Cotton gravity  was recently proposed by Harada \cite{Harada}. We showed that the field equation can be recast as Einstein equations, with the freedom of
a Codazzi tensor. As such, they are second order in the derivatives of the metric tensor \cite{Codazzi}.

This is the ouline of the paper. In \S2 we discuss static space-times in general, the Ricci and the Weyl tensors. It is
partly based on our study of the larger family of doubly-warped space-times \cite{DWST}. Useful equations are collected in Appendix 1.\\
In \S 3 we introduce the Faraday tensor and prove necessary and sufficient conditions on the scalars $\mathbb E$ and
$\mathbb B$ (the proof is in Appendix 2). The general discussion of Einstein gravity coupled to linear (LE)  and non-linear electrodynamics (NLE) is in \S4. An interesting form of the Ricci tensor is obtained, with conclusions about the curvature 
space-time and space scalars $R$ and $R^\star$. \S5 discusses the anisotropic fluid source, concluding that the energy
density is proportional to $R^\star \ge 0$.\\
In \S6 we discuss spherical symmetry, where the full form of the Ricci, Weyl, Cotton and Bach tensors are obtained, 
as combinations of the basic tensors $g_{jk}$, $u_ju_k$ and $\dot u_j \dot u_k$ with coefficients that are linear or
at most quadratic in $B(r)$, $B'(r)$ and $B''(r)$. Simple conditions yield the early static metrics. 
In \S7 we consider the Einstein gravity, where the metrics are solutions of field equations. Pure dust or perfect fluid solutions are not possible, unless $p=-\mu$. We then discuss LE and NLE, with some identities that allow for reconstructing the Lagrangian $\mathscr L(F)$ from $B(r)$ (actually from a
coefficient of the Ricci tensor).\\
NLE coupled to $f(R)$ gravity is presented in \S8, with immediate recognition of the known property that $f_R(r)$ is
constrained to be linear in $r$. This allows the integration of one field equation in presence of point charges. Since
only the solution appears in previous papers, we offer its deduction in Appendix 3.\\
In Cotton gravity (\S9), after a brief presentation of our interpretation as an Einstein equation, we show that the vacuum solution by Harada is also a solution of the Einstein theory with an anisotropic energy-momentum. We then
present two solutions: with perfect fluid and LE. Finally, in \S10 we turn to Conformal gravity. We recall the vacuum solution by Mannheim and Kazanas, and present a solution in LE. We end with the conclusions.

In this paper the static four-dimensional Lorentzian spacetimes have signature $(-,+++)$. A dot denotes the action of $u^k\nabla_k$.

\section{Static space-times}
\begin{center}
{\em The velocity is eigenvector of the Ricci tensor, the Electric tensor\\ evaluates the Weyl tensor, the 
form of the Ricci tensor is obtained.}
\end{center}
\noindent
There are various characterisations of static spacetimes: \\
$\bullet$ The existence of a time-like vector field $u_k$  (named velocity) that is normalized,  $u^ku_k=-1$, with gradient
\begin{align}
\nabla_j u_k = -u_j \dot u_k  \label{nablau}
\end{align}
such that the `acceleration' $\dot u_k = u^j\nabla_j u_k$ is a closed vector field \cite{Stephani}:
\begin{align}
\nabla_j \dot u_k = \nabla_k \dot u_j  \label{closed}
\end{align}
The acceleration is spacelike ($u^k\dot u_k=0$) with normalization $\eta =\dot u^p\dot u_p>0$.
Contraction of \eqref{closed} with $u^j$ shows that $\ddot u_k = u^j\nabla_k \dot u_j = - \dot u^j \nabla_k u_j = \dot u^j u_k \dot u_j = \eta u_k$.\\ 
%
$\bullet$ The existence of coordinates $(t,{\bf x})$ where the metric tensor has the static form 
$$ ds^2 = -B({\bf x})dt^2 + g^\star_{\mu\nu} ({\bf x}) dx^\mu dx^\nu $$
In this frame: $u_k=(-\sqrt B, {\bf 0})$, $\dot u_k =(0, \partial_\mu \log\sqrt B)$.\\
$\bullet$ The existence of a time-like hypersurface orthogonal Killing vector: $\nabla_i\xi_j+\nabla_j\xi_i=0$. The vector is $\xi_j=u_j\sqrt B$, and $\dot u_j =\nabla_j \log \sqrt B$.
%

\begin{prop}\label{P1}
The velocity is an eigenvector of the Ricci tensor,
\begin{align}
R_{jk} u^k = - (\nabla_p \dot u^p) u_j \label{EIGV}
\end{align}
and it is Riemann compatible:
\begin{align}
(u_i R_{jklm}+u_j R_{kilm} + u_k R_{ijlm}) u^m =0.
\end{align}
\begin{proof} 
$R_{jklm}u^m = (\nabla_j\nabla_k -\nabla_k\nabla_j)u_l = (u_j\dot u_k -\dot u_j u_k)\dot u_l +
u_j\nabla_k \dot u_l - u_k\nabla_j \dot u_k$. 
Contraction with $g^{jl}$ gives property \eqref{EIGV}. Multiplication by $u_i$ and cyclic sum gives compatibility.
\end{proof}
\end{prop}

As expected, the ``time derivative'' of geometric invariants is zero: 
\begin{prop} Let $\eta = \dot u^k\dot u_k$ and $R=g^{jk} R_{jk}$,
\begin{align}
\dot \eta =0,\quad \dot R=0, \quad u^k\nabla_k (\nabla_p\dot u^p)=0
\end{align}
\begin{proof}
1) $\dot \eta = u^k\nabla_k (\dot u^p \dot u_p) = 2\ddot u^p u_p = 2\eta u^p\dot u_p =0$. 2) $u^k\nabla_k \nabla_p\dot u^p  = u^k R_{kp}{}^p{}_m \dot u^m + u^k\nabla_p\nabla_k \dot u^p = u^k R_{kp}\dot u^p + \nabla_p(\ddot u^p) - (\nabla_p u^k)(\nabla_k\dot u^p) = \nabla_p (\eta u^p) + u^p\dot u^k  \nabla_p \dot u_k = \dot \eta + \frac{1}{2}\dot \eta = 0$. 
3) Eq.\eqref{EIGV} and $\nabla_j R^j{}_k = \frac{1}{2}\nabla_k R$ give $\dot R = -2 u^j\nabla_j (\nabla_p\dot u^p)=0$.
\end{proof}
\end{prop}
%
Being Riemann compatible, the velocity is also ``Weyl compatible" (Theorem 2.1 in \cite{WCOMP}):
$(u_i C_{jklm}+u_j C_{kilm} + u_k C_{ijlm}) u^m =0$. The Weyl tensor is
\begin{align*}
C_{jklm} = R_{jklm} + \frac{1}{2}(g_{jm} R_{kl} - g_{km} R_{jl} + g_{kl}R_{jm} -
g_{jl} R_{km}) - \frac{R}{6} \, ( g_{jm}g_{kl}-
g_{km}g_{jl} ).
\end{align*}
The contraction $E_{kl} = u^j C_{jklm} u^m$ is the Electric tensor; it is symmetric, traceless and $E_{jk}u^k=0$. 
Weyl compatibility is equivalent to the relation 
$$C_{jklm}u^m = u_k E_{jl} - u_j E_{kl} $$
The explicit evaluation of the Electric tensor gives an identity with the Ricci tensor:
\begin{align}
E_{kl} =  - \frac{1}{2} R_{kl}  +\frac{1}{2}(\nabla_p\dot u^p) (g_{kl} + 2u_l u_k)   + \frac{1}{6} R (g_{kl}+
u_k u_l ) + u^jR_{jklm} u^m \label{EandRICCI}
\end{align}
where $u^j R_{jklm}u^m = -\dot u_k \dot u_l - \nabla_k \dot u_l - \eta u_k u_l$ (see Prop.\ref{P1}).

\begin{prop}[Weyl tensor] \quad\\
In a four-dimensional static spacetime the Weyl tensor is solely determined by the electric tensor:
\begin{align}
C_{jklm} =& (g_{kl}+2 u_k u_l )E_{jm}  -(g_{jl}+2u_ju_l )E_{km} \label{WEYLT}\\
&+ (g_{jm} +2u_j u_m) E_{kl} - (g_{km}+2u_k u_m) E_{jl}   \nonumber
\end{align}
and $C_{jklm}C^{jklm} = 8 E^{kl}E_{kl}$.
\begin{proof}
In $n=4$ the following identity by Lovelock holds \cite{LovRund}:
$0= g_{ir} C_{jklm} + g_{jr} C_{kilm}+ g_{kr} C_{ijlm} +g_{im} C_{jkrl} + g_{jm} C_{kirl}+ g_{km} C_{ijrl} 
+g_{il} C_{jkmr} + g_{jl} C_{kimr}+ g_{kl} C_{ijmr} $.
The contraction with $u^i u^r$ and Weyl compatibility give the Weyl tensor.
\end{proof}
\end{prop}
At each point we choose a basis of vectors formed
by $u_i$ and three orthonormal space-like vectors $\frac{1}{\sqrt\eta}\dot u_i$, $y_i$, $z_i$:
\begin{align}
g_{ij} = -u_i u_j + \frac{\dot u_i \dot u_j}{\eta}   +y_i y_j + z_i z_j \label{IDENTITY}
\end{align}
\begin{lem}\label{LEMMAYZ}
In a $n=4$ static space-time:
\begin{align}
&\nabla_k y_j = - Y_k \dot u_j -  \Omega_k z_j \label{nablay}\\
&\nabla_k z_j = -Z_k\dot u_j +\Omega_k y_j \label{nablaz}\\
&\nabla_k \dot u_j = -\eta u_k u_j +\tfrac{1}{2\eta}\dot u_j\nabla_k\eta+\eta (y_j Y_k  + z_j Z_k) \label{NABDOTU}
\end{align}
where $Y_k =\frac{1}{\eta} y^p\nabla_k \dot u_p$, $Z_k =\frac{1}{\eta} z^m\nabla_k \dot u_m$ and $\Omega_k=
y^p\nabla_k z_p$. It is also: 
\begin{gather}
u^k Y_k=0,\quad u^kZ_k=0, \quad y^k Z_k = z^k Y_k \nonumber \\
\eta ( y^k Y_k + z^k Z_k ) =  - \eta + \nabla^p\dot u_p   - \frac{\dot u^p \nabla_p\eta}{2\eta} \label{yYzZ}
\end{gather}
\begin{proof}
The gradient of \eqref{IDENTITY} is: $ 0 = u_k (u_i \dot u_j + \dot u_i u_j) 
+\frac{1}{\eta} (\dot u_i \nabla_k \dot u_j  + \dot u_j \nabla_k \dot u_i) 
+y_i \nabla_k y_j +y_j \nabla_k y_i + z_i \nabla_k z_j+ z_j \nabla_k z_i $. The contractions with $y^i$ or $z^i$ give the first 
two relations. While contracting with $\dot u^i$ note that $\dot u^i\nabla_k y_i = -y^i\nabla_k \dot u_i =-\eta Y_k$, and
$\dot u^i\nabla_k z_i =-\eta Z_k$.\\
Similarly: $u^kY_k= \frac{1}{\eta} y^j \ddot u_j  = y^ju_j=0$ and $u^j Z_j=0$, and $y^kZ_k=z^kY_k$. Finally, \eqref{yYzZ}
results from the contraction $g^{jk}$ of \eqref{NABDOTU}.
\end{proof}
\end{lem}

With this choice of basis vectors, the Ricci tensor \eqref{EandRICCI} is:
\begin{align}
 R_{kl} =&u_k u_l \left[ \frac{R}{3} +2\nabla_p \dot u^p\right ] + g_{kl}\left [ \frac{R}{3} +\nabla_p \dot u^p\right ] -2\dot u_k \dot u_l -2E_{kl}
 \label{RICCISTATIC}\\
 & -\frac{1}{\eta} \dot u_l \nabla_k \eta -2\eta (Y_k y_l + Z_k z_l).\nonumber
\end{align}
In static space-times the curvature scalar $R$ and the space curvature scalar $R^\star $ are 
related by the identity (see \cite{DWST} eq. 34):
\begin{align}
R=R^\star -2\nabla_p \dot u^p. \label{RRSTAR} 
\end{align}

\section{The Faraday tensor in static space-times}
\begin{center}
{\em The conditions for a Faraday tensor and the conserved current are given.}
\end{center}

The antisymmetric tensors $u_i\dot u_j - u_j \dot u_i$ and $y_i z_j - y_jz_i$ are ``time independent'':
$ u^k\nabla_k (u_i\dot u_j - u_j \dot u_i)=0$ and $u^k\nabla_k (y_i z_j - y_jz_i) =0 $. 
The other antisymmetric combinations of the basis vectors do not share this property. 

Therefore, we consider the following antisymmetric tensor 
\begin{align}
F_{jk} = \frac{\mathbb E}{\sqrt\eta} (u_j \dot u_k - \dot u_j u_k) + {\mathbb B} (y_j z_k - y_k z_j) \label{FTENSOR}
\end{align}
where $\mathbb E$ and $\mathbb B$ are scalar fields with $\dot {\mathbb E}=\dot {\mathbb B}=0$. Since $\dot\eta=0$, it is 
$\dot F_{jk}=0$.\\ 
$F_{jk}$  is a Faraday tensor if:
\begin{align}
\nabla_j F_{kl} + \nabla_k F_{lj} + \nabla_l F_{jk} = 0. \label{FARADAY}
\end{align}

\begin{thrm} [The Faraday tensor]\label{THFAR}\quad\\
In a static space-time, the tensor \eqref{FTENSOR} with $\dot {\mathbb E}=\dot {\mathbb B}=0$ is Faraday if and only if
\begin{align}
&\nabla_k \frac{\mathbb E}{\sqrt\eta} = \varkappa \dot u_k, \label{FARADAYCONDQ}\\
&\dot u^k \nabla_k \mathbb B = \mathbb B \left[ \eta - \nabla^k\dot u_k   + \frac{\dot u^k \nabla_k\eta}{2\eta} \right ].  \label{FARADAYCONDP}
\end{align}
Then it is $\dot\varkappa =0$.
\begin{proof} 
See Appendix 2.
\rem{SEEAPPENDIX
With \eqref{FTENSOR} and Lemma \ref{LEMMAYZ}:
\begin{align}
 \nabla_i F_{jk} 
=&
(\nabla_i \frac{\mathbb E}{\sqrt\eta})(u_j \dot u_k - \dot u_j u_k) + (\nabla_i P)(y_j z_k - y_k z_j) 
+\frac{\mathbb E}{\sqrt \eta}(u_j\nabla_i \dot u_k - u_k \nabla_j \dot u_i) \nonumber \\
&+P(- Y_i \dot u_j z_k - Z_i y_j \dot u_k  + Y_i z_j\dot u_k + Z_i\dot u_j y_k)  \label{nablaF}
\end{align} 
The cyclic sum is:
\begin{align*}
&(\nabla_i \frac{\mathbb E}{\sqrt \eta})(u_j \dot u_k - \dot u_j u_k)+(\nabla_j \frac{\mathbb E}{\sqrt\eta})(u_k\dot u_i - \dot u_k u_i)+ (\nabla_k \frac{\mathbb E}{\sqrt \eta})(u_i \dot u_j - \dot u_i u_j)\\
&+ (\nabla_i \mathbb B)(y_j z_k - y_k z_j)+(\nabla_j \mathbb B)(y_k z_i - y_i z_k)+(\nabla_k \mathbb B)(y_i z_j - y_j z_i)\\
&+\frac{\mathbb E}{\sqrt \eta}(u_j\nabla_i \dot u_k - u_k \nabla_i \dot u_j +u_k\nabla_j \dot u_ i- u_i \nabla_j \dot u_k +u_i\nabla_k \dot u_j - u_j \nabla_k \dot u_i )\\
&+ \mathbb B Y_i (z_j\dot u_k -z_k \dot u_j) - \mathbb B Z_i (y_j \dot u_k - y_k\dot u_j) +
 \mathbb B Y_j (z_k\dot u_i -z_i\dot u_k) - \mathbb B Z_j (y_k \dot u_i -y_i\dot u_k) \\
&+ \mathbb B Y_k (z_i \dot u_j -z_j\dot u_i) - \mathbb B Z_k (y_i \dot u_j  - y_j\dot u_i) 
\end{align*}
The third line is zero because the acceleration is closed.
For the cyclic sum to be zero, all contractions with vectors must be zero, and give conditions. Contraction with $u^i$ gives:
$(\nabla_j \frac{\mathbb E}{\sqrt\eta}) \dot u_k - (\nabla_k \frac{\mathbb E}{\sqrt \eta}) \dot u_j = 0$ with solution 
$$\nabla_j (\frac{\mathbb E}{\sqrt\eta} )= \varkappa \dot u_j$$
With this result the ciclic condition simplifies: 
\begin{align*}
& (\nabla_i \mathbb B)(y_j z_k - y_k z_j)+(\nabla_j \mathbb B)(y_k z_i - y_i z_k)+(\nabla_k \mathbb B)(y_i z_j - y_j z_i)\\
&+ \mathbb BY_i (z_j\dot u_k -z_k \dot u_j) - \mathbb B Z_i (y_j \dot u_k - y_k\dot u_j) +
 \mathbb B Y_j (z_k\dot u_i -z_i\dot u_k) - \mathbb B Z_j (y_k \dot u_i -y_i\dot u_k) \\
&+ \mathbb B Y_k (z_i \dot u_j -z_j\dot u_i) - \mathbb B Z_k (y_i \dot u_j  - y_j\dot u_i)=0. 
\end{align*}
Contraction with $y^i$:
 \begin{align*}
& (y^i\nabla_i \mathbb B )(y_j z_k - y_k z_j) - (\nabla_j \mathbb B) z_k+ (\nabla_k \mathbb B) z_j \\
&+ \mathbb B y^iY_i (z_j\dot u_k -z_k \dot u_j) - \mathbb B y^iZ_i (y_j \dot u_k - y_k\dot u_j) +\mathbb B (Z_j \dot u_k - Z_k \dot u_j )  =0.
\end{align*}
A further contraction with $z^j$ gives:
$\nabla_k \mathbb B =  (y^i\nabla_i \mathbb B) y_k  +(z^j\nabla_j \mathbb B) z_k - \mathbb B (y^jY_j + z^jZ_j )\dot u_k $ i.e.  
$\dot u^k \nabla_k \mathbb B = - \eta \mathbb B (y^rY_r +z^r Z_r )$.
The right-hand-side is evaluated in Lemma \ref{LEMMAYZ} and gives the second condition.

Using the form of $\nabla_j \mathbb B$, the cyclic condition becomes 
\begin{align*}
& -(y^rY_r + z^rZ_r ) [\dot u_i  (y_j z_k - y_k z_j)+ \dot u_j (y_k z_i - y_i z_k)+\dot u_k (y_i z_j - y_j z_i) ]\\
&+ Y_i (z_j\dot u_k -z_k \dot u_j) - Z_i (y_j \dot u_k - y_k\dot u_j) +
 Y_j (z_k\dot u_i -z_i\dot u_k) - Z_j (y_k \dot u_i -y_i\dot u_k) \\
&+ Y_k (z_i \dot u_j -z_j\dot u_i) - Z_k (y_i \dot u_j  - y_j\dot u_i) =0.
\end{align*}
The contractions with $\dot u^i$, $y^i$ or $z^i$ or with the metric tensor are trivial. Indeed it is satisfied by the generic 
expansions $Y_i = ay_i + b z_i + c \dot u_i$ and $Z_i = a' y_i + b' z_i + c' \dot u_i$.

$\dot\varkappa = u^k\nabla_k (\frac{1}{\eta}\dot u^j \nabla_j \frac{\mathbb E}{\sqrt\eta})$. Use $\dot \eta=0$, $\ddot u^j =\eta u^j$ and $u^j\nabla_j \frac{\mathbb E}{\sqrt\eta}=0$. Then
$\dot\varkappa =\frac{1}{\eta}  \dot u^j u^k\nabla_k \nabla_j \frac{\mathbb E}{\sqrt\eta} = \frac{1}{\eta}  \dot u^j u^k\nabla_j \nabla_k \frac{\mathbb E}{\sqrt\eta} $. Now: $$u^k\nabla_j \nabla_k \frac{\mathbb E}{\sqrt\eta} =
\nabla_j (u^k\nabla_k \frac{\mathbb E}{\sqrt\eta}) - (\nabla_j u^k)\nabla_k \frac{\mathbb E}{\sqrt\eta} = \nabla_j (\varkappa  u^k \dot u_k) + u_j \dot u^k (\varkappa \dot u_k) = u_j\eta\varkappa $$
Then: $\dot \varkappa = \dot u^j u_j \varkappa =0 $.
ENDAPPENDIX}
\end{proof}
\end{thrm}

\noindent
The vector field $J_k =\nabla_j F^j{}_k $ is a conserved current $\nabla_k J^k=0$. 
\begin{prop}[The current]
\begin{align}
 &J^k = J u^k + (z^k y^m - y^kz^m) (\nabla_m \mathbb B - \mathbb B \frac{\nabla_m \eta}{2\eta})
 \label{CURRENT}\\
 &J= -\eta \varkappa +\mathbb E\sqrt \eta -\frac{\mathbb E}{\sqrt\eta}\nabla_j \dot u^j
\end{align} 
\begin{proof}
Eq.\eqref{nablaF} gives $ J_k = -\eta \varkappa  u_k +\frac{\mathbb E}{\sqrt \eta}(\ddot u_k - u_k \nabla_i \dot u^i)
 + (y^j\nabla_j \mathbb B - \mathbb B\dot u^j Y_j) z_k   - (z^j\nabla_j \mathbb B - \mathbb B \dot u^j Z_j)y_k +
 \mathbb B(Y_j z^j - Z_j y^j) \dot u_k $.
The last term is zero and $\ddot u_k =\eta u_k$. It is also
$ \dot u^j Y_j = \frac{1}{\eta} y^m \dot u^j \nabla_j \dot u_m =\frac{1}{\eta} y^m \dot u^j \nabla_m \dot u_j =\frac{1}{2\eta} y^m \nabla_m \eta $, and $\dot u^j Z_k = \frac{1}{2\eta} z^m \nabla_m \eta $.
The current is obtained, and is orthogonal to $\dot u_k$. 
\end{proof}
\end{prop}

%
%
%
\section{Linear / non-linear electrodynamics in static Einstein gravity}
\begin{center}
{\em The equations of the Einstein - LE and NLE theory are discussed.}
\end{center}

The Einstein equations of gravity coupled to an electromagnetic field descend from the action (see \cite{Ayon00})
$$ S=\frac{1}{2}\int d^4x \sqrt{-g} \left [ R - 4 \mathscr L ( F) \right ] $$
where $\mathscr L$ is a scalar function of the squared Faraday tensor $F=\tfrac{1}{4} F_{jk}F^{jk} $.
In linear electrodynamics $\mathscr L(F)=F$.\\
The vanishing of the variations of the action in the metric tensor and in the vector potential ($F_{jk}=\nabla_j A_k - \nabla_k A_j$) 
respectively give:
\begin{align}
&R_{jk} -\tfrac{1}{2} g_{jk} R= 2 \mathscr L_F(F)  F_{jm} F_k{}^m - 2 g_{jk} \mathscr L (F) \label{EINSTEINEQ}\\
&\nabla_j (\mathscr L_F(F) F^{jk}) =0 \label{CURRENTEQ}
\end{align}
where $\mathscr L_F = d\mathscr L/dF$. The right-hand-side of eq.\eqref{EINSTEINEQ} is the energy-momentum tensor $T_{jk}^{nlin} $ of non-linear electrodynamics.\\ 
In the static setting with $F_{jk}$ given by \eqref{FTENSOR}, it is
$F = \tfrac{1}{2}(\mathbb B^2-\mathbb E^2)$ and 
\begin{align}
T_{jk}^{nlin} = 2 (\mathbb E^2+\mathbb B^2) \left [ u_j u_k  -\frac{\dot u_j \dot u_k}{\eta} \right ] \mathscr L_F (F)
+ 2 g_{jk} [\mathbb B^2 \mathscr L_F(F) - \mathscr L(F)].  \label{NONLINENMOMTENS}
\end{align}
Note the disappearance of the space-like vectors $y_j $ and $z_j$.\\
In the linear case the tensor is traceless:
\begin{align}
T_{jk}^{lin} = 2 (\mathbb E^2+\mathbb B^2) \left [ u_j u_k +\frac{1}{2}g_{jk} -\frac{\dot u_j \dot u_k}{\eta} \right ]. 
 \label{LINENMOMTENS}
\end{align}
The contractions of the Einstein equation \eqref{EINSTEINEQ} with $u^j$ and $g^{jk}$ give two interesting relations between the geometry and the scalars of electrodynamics:
\begin{align}
&\mathbb E^2 \mathscr L_F (F) + \mathscr L (F)=\tfrac{1}{2}\nabla_p\dot u^p + \tfrac{1}{4}R \label{FFF1}\\
&F \mathscr L_F (F) - \mathscr L(F) = - \tfrac{1}{8} R \label{FFF2}
\end{align}
These are immediate consequences:
\begin{prop} {\quad}\\
1) $R=0$ if and only if $\mathscr L(F)=cF$ (we take $c=1$).\\ 
2) In linear electrodynamics: $\mathbb B^2+\mathbb E^2 = \nabla_p\dot u^p $.
\end{prop}

The sum of \eqref{FFF1} and \eqref{FFF2} is: $(\mathbb B^2+\mathbb E^2)\mathscr L_F=\nabla_p\dot u^p +\tfrac{1}{4}R$. The Einstein equation of non-linear electrodynamics 
becomes a geometric prescription for the Ricci tensor, which acquires a form much simpler than the general one in static space-times \eqref{RICCISTATIC}:
\begin{align}
 R_{jk} = (R^\star -\frac{1}{2}R)  \left [ u_j u_k  +\frac{1}{2} g_{jk}-\frac{\dot u_j \dot u_k}{\eta} \right ] 
+ \frac{1}{4} g_{jk}  R 
\end{align}
$u^k$ and $\dot u^k$ are eigenvectors of the Ricci tensor with eigenvalue $- (\nabla_p \dot u^p) $, while $y^k$ and $z^k$ are eigenvectors with eigenvalue $\frac{R}{2}+\nabla_p\dot u^p$ (in the linear case: $R=0$).

The second field equation \eqref{CURRENTEQ} describes the current. It is equivalent to the following three equations
\begin{align}
& \nabla_j  \left [\tfrac{\dot u^j}{\sqrt\eta} \mathbb E \mathscr L_F\right ] = \sqrt \eta \left [ \mathbb E\mathscr L_F\right ]  \label{NLE1}\\
&\nabla_j [y^j \mathbb B \mathscr L_F] = - (z^j\Omega_j) [ \mathbb B \mathscr L_F] \label{NLE2}\\
&\nabla_j [z^j \mathbb B \mathscr L_F ] = (y^j\Omega_j) [\mathbb B \mathscr L_F ]  \label{NLE3} 
\end{align}
\begin{proof}
The expression \eqref{FTENSOR} of the Faraday tensor is placed in \eqref{CURRENTEQ}:
\begin{align*}
0=&\nabla_j [\tfrac{\mathbb E}{\sqrt\eta} \mathscr L_F (u^j \dot u_k - u_k \dot u^j)] + \nabla_j [\mathbb B \mathscr L_F (y^j z_k - y_k z^j)] \\
=&\dot u_k u^j\nabla_j [\tfrac{\mathbb E}{\sqrt\eta} \mathscr L_F ] - u_k \dot u^j \nabla_j [\tfrac{\mathbb E}{\sqrt\eta} \mathscr L_F ] +
 [\tfrac{\mathbb E}{\sqrt\eta} \mathscr L_F ] (\ddot u_k - u_k \nabla_j \dot u^j) \\
 &+ z_k y^j\nabla_j [\mathbb B \mathscr L_F ] - y_k z^j  \nabla_j[\mathbb B \mathscr L_F ] +
 [\mathbb B \mathscr L_F ] (z_k\nabla_j y^j + y^j\nabla_j z_k - y_k \nabla_j z^j - z^j \nabla_j y_k) \\
 =& - u_k \dot u^j \nabla_j [\tfrac{\mathbb E}{\sqrt\eta} \mathscr L_F ] + u_k
 [\tfrac{\mathbb E}{\sqrt\eta} \mathscr L_F ] (\eta - \nabla_j \dot u^j) + z_k y^j\nabla_j [\mathbb B \mathscr L_F ] - y_k z^j \nabla_j
  [\mathbb B \mathscr L_F ] \\
 &+[\mathbb B \mathscr L_F ] (z_k\nabla_j y^j - y^jZ_j\dot u_k +y^j\Omega_j y_k - y_k \nabla_j z^j + z^j Y_j \dot u_k 
 +z^j\Omega_j  z_k) 
 \end{align*}
The coefficient of $\dot u_k$ is proportional to $(z^j Y_j -y^j Z_j)=0$. The vector equation gives three conditions for the coefficients of the components along $u_k$, $z_k$ and $y_k$.
\end{proof}
An extension with $\mathscr L(F,{}^*F)$, where the invariant scalar ${}^*F$ is built with the dual Faraday tensor, is studied 
by Bokuli\'c et al. \cite{Bokulic}.

\section{Anisotropic perfect fluid in static Einstein gravity}
\begin{center}
{\em Absence of convective term. Positive energy means positive space-curvature scalar.}
\end{center}
The Einstein equation for a static anisotropic fluid with velocity $u_i$ is:
$$ R_{jk} - \frac{1}{2} g_{jk} R = (p+\mu) u_j u_k + p g_{jk} + \Pi_{jk} $$
where $\Pi_{jk}$ is the stress-tensor (traceless and $\Pi_{jk} u^k=0$), $p$ is the effective pressure and $\mu$ is the energy density. A convective term $(u_j q_k + u_k q_j)$ is forbidden in static space-times as it would violate eq.\eqref{EIGV}.\\
By the general property $R_{jk}u^k = (-\nabla_p \dot u^p) u_j$ the contraction of the Einstein equation with $u^k$ gives $\nabla_p\dot u^p + \frac{R}{2}= \mu $. Now use \eqref{RRSTAR} and obtain the simple relation:
\begin{align}
\mu = \frac{1}{2} R^\star
\end{align}
The trace and the previous equation give the pressure:
\begin{align}
3p= \frac{1}{2} R^\star -R
\end{align}
\begin{remark}
 In general, in a static space-time the Einstein equations relate the positive energy constraint to the space curvature
 scalar:
 $$T_{ij} u^i u^j = R_{jk}u^ju^k +\tfrac{1}{2}R = \nabla_p \dot u^p +\tfrac{1}{2} R =\tfrac{1}{2}R^\star \ge 0 $$ 
In spherical symmetry (see Appendix 1): $R^\star = 2 (\frac{1-B}{r^2} +\frac{B'}{r})$. The condition becomes
\begin{align}
\frac{d}{dr} \frac{B(r)-1}{r} \ge 0 
\end{align}
 \end{remark}

\section{Spherical symmetry}\label{SPHERICAL}
\begin{center}
{\em Expressions of the Ricci, Weyl, Cotton and Bach tensors in terms of\\ 
$u_ju_k$, $g_{jk}$ and $\dot u_j \dot u_k$. 
Natural constraints give notorious metrics.\\ The magnetic part of the Faraday tensor is a monopole.}
\end{center}

The majority of static spherical metrics discussed in the literature depend on a single scale function $B(r)>0$:
\begin{align}
 ds^2 = - B(r) dt^2 + \frac{dr^2}{B(r)} + r^2 ( d\theta^2 + \sin^2 \theta\, d\phi^2 ). \label{metric}
 \end{align}
 In coordinates $(t,r,\theta,\phi)$, the acceleration is the radial vector $\dot u_k=(0,\frac{B'}{2B},0,0)$, where a prime is a derivative in $r$. \\
If $X(r)$ is a scalar function, its gradient is parallel to $\dot u_j$:
\begin{align}
 \nabla_j X =  \frac{\dot u_j\dot u^k}{\eta}\nabla_k X= \dot u_j \frac{2B}{B'} X'  \label{GRAD}
 \end{align}
The following scalars are obtained from expressions valid for the broader class of spherical doubly-warped space-times
(see Appendix 1 or equations 51, 40 and 53 in \cite{DWST}):
\begin{align}
\eta = \frac{1}{4} \frac{B'^2}{B} , \qquad \frac{\dot u^p\nabla_p \eta}{\eta}    =  B'' - \frac{B'^2}{2B}, \qquad  \nabla_p\dot u^p = \frac{B''}{2} + \frac{B'}{r}. 
\label{various}
 \end{align} 
A key quantity is the following:
 \begin{prop}\label{Propnabladotu}
 \begin{align}
\nabla_j \dot u_k = \left[ -\eta + \frac{B'}{2r}\right ]  u_j u_k  + \frac{B'}{2r} g_{jk} +  
\left[ \frac{B''}{2} - \frac{B'}{2r} -\eta \right ] \frac{\dot u_j\dot u_k}{\eta}. \label{nabladot}
\end{align}
 \begin{proof}
In spherical coordinates, with $y^j=(0,0,r,0)$ and $z^j=(0,0,0,r\sin\theta)$ one evaluates
 \begin{align*}
 Y_j = \frac{2B}{rB'}y_j, \qquad Z_j = \frac{2B}{rB'}z_j
 \end{align*}
 Then $Y_j y_k + Z_j z_k = \frac{2B}{rB'} (g_{jk} + u_j u_k -\frac{ \dot u_j\dot u_k}{\eta} )$. The static expression \eqref{NABDOTU} becomes
 \begin{align*}
 \nabla_j \dot u_k =& -\eta u_j u_k +\frac{\dot u_j \dot u_k}{2\eta^2}\dot u^p\nabla_p\eta+\eta  \frac{2B}{rB'} (g_{jk} + u_j u_k -\frac{ \dot u_j\dot u_k}{\eta} )\\
 &=u_j u_k \left[ -\eta + \frac{B'}{2r}\right ] + g_{jk} \frac{B'}{2r} + \frac{\dot u_j\dot u_k}{\eta} \left[ \frac{1}{2\eta}\dot u^p\nabla_p \eta  -  \frac{B'}{2r} \right ]
 \end{align*}
 With insertions of the scalars \eqref{various}, the result is obtained.
\end{proof}
 \end{prop}
 
We now obtain the covariant expressions of the Ricci, the Weyl, the Cotton and the Bach tensors. They will appear as
combinations of the tensors $u_j u_k$, $g_{jk}$ and $\dot u_j\dot u_k$,
with coefficients that are scalar functions of $r$. 

\begin{prop}[\underline{The Ricci tensor}]
\begin{align}
&R_{jk} = g_{jk} \frac{R(r)}{4}  + \left[ u_j u_k +\frac{1}{2}g_{jk} -\frac{\dot u_j \dot u_k}{\eta} \right ] A(r)  \label{RicciT} \\
& A(r) = \frac{1-B}{r^2} +\frac{B''}{2} \nonumber
\end{align}
\begin{proof} 
Insert the expression for  $Y_ky_l +Z_kz_l$ evaluated in Prop.\ref{Propnabladotu} in eq.\eqref{RICCISTATIC}:
\begin{align*}
 R_{kl} =&u_k u_l \left[ \tfrac{R}{3} +2\nabla_p \dot u^p\right ] + g_{kl}\left [ \tfrac{R}{3} +\nabla_p \dot u^p\right ] -2\dot u_k \dot u_l -2E_{kl}\\
 & -\tfrac{\dot u_k \dot u_l}{\eta} \tfrac{\dot u^p \nabla_p\eta}{\eta} -2\tfrac{B'^2}{4B}  \tfrac{2B}{rB'} (g_{kl} + u_k u_l -\tfrac{ \dot u_k\dot u_l}{\eta} )\nonumber\\
 =&u_k u_l \left[ \tfrac{R}{3} +2\nabla_p \dot u^p -\tfrac{B'}{r}\right ] + g_{kl}\left [ \tfrac{R}{3} +\nabla_p \dot u^p  -\tfrac{B'}{r}\right ]  \\
 & -\tfrac{\dot u_k \dot u_l}{\eta} \left [2\eta + \tfrac{\dot u^l \nabla_l\eta}{\eta} -\tfrac{B'}{r}  \right ]  -2E_{kl}\nonumber
\end{align*}
The spherical scalars $\eta $, $\dot u^p\nabla_p\eta $ and $\nabla_p\dot u^p$ are given in eq.\eqref{various}. 
The electric tensor and the curvature scalar are obtained from equations \eqref{ELECTRIC2} and \eqref{SCALARR} in Appendix 1:
\begin{align}
&E_{kl} = E(r)  \left [\frac{\dot u_k \dot u_l}{\eta} - \frac{u_k u_l+g_{kl}}{3}\right ]\\
&E(r)  = \frac{1}{2}\left[ \frac{1-B}{r^2} + \frac{B'}{r} - \frac{B''}{2}\right ] \nonumber\\
&R(r)= 2\frac{1-B}{r^2} - 4\frac{B'}{r} - B''
\end{align}
The Ricci tensor is then written as a sum with a trace-less term.
\end{proof}
\end{prop}
\begin{prop}[\underline{The Weyl tensor}]\quad\\
With the expression of the electric tensor, the static Weyl tensor \eqref{WEYLT} becomes:
\begin{align}
C_{jklm} = E(r) [& (g_{kl}+2 u_k u_l )\frac{\dot u_j \dot u_m}{\eta}  -(g_{jl}+2u_ju_l )\frac{\dot u_k \dot u_m}{\eta} 
 + (g_{jm} +2u_j u_m) \frac{\dot u_k \dot u_l}{\eta} \nonumber \\
 & - (g_{km}+2u_k u_m) \frac{\dot u_j \dot u_l}{\eta}  
- u_k u_l  g_{jm} +u_ju_l g_{km}  - u_j u_m g_{kl} + u_k u_m g_{jl} \nonumber\\
&-\frac{2}{3} (g_{kl} g_{jm} - g_{jl}g_{km}) ]. \label{WEYL}
 \end{align}
\end{prop}
\noindent
The Riemann tensor can then be obtained.

\begin{prop} {\rm \underline{The Cotton tensor} (Cotton, 1899, \cite{Cotton1899})} \\ 
The Cotton tensor $C_{jkl} = \nabla_j R_{kl} - \nabla_k R_{jl} - \frac{1}{6}(g_{kl} \nabla_j R - g_{jl} \nabla_k R) $ 
is proportional to $\nabla_m C_{jkl}{}^m $. Here it is
\begin{align}
C_{jkl} = \frac{B'}{2\eta} \left ( A' +\frac{A}{r}\right ) \left [ \dot u_j (u_k u_l + \frac{1}{3} g_{kl} ) - (u_j u_l +\frac{1}{3} g_{jl} )\dot u_k  \right ]. \label{Cotton}
\end{align}
\begin{proof}
The evaluation is rather long. Let us specify some building steps. With \eqref{nabladot} we obtain:
\begin{align}
 \nabla_j \frac{\dot u_k \dot u_l}{\eta} = - u_j (u_k \dot u_l +u_l \dot u_k) + \frac{B'}{2r\eta}(h_{jk}\dot u_l +h_{jl}\dot u_k - \frac{2}{\eta}
\dot u_j \dot u_k \dot u_l ) \label{NABLAXXX}
\end{align}
Next, with the spherical static Ricci tensor \eqref{RicciT}:
\begin{align*}
\nabla_j (R_{kl} -\frac{R}{6}g_{kl} )=& -A(r)\frac{B'}{2r\eta}(h_{jk}\dot u_l +h_{jl}\dot u_k - \frac{2}{\eta}\dot u_j \dot u_k \dot u_l ) \\
&+ \frac{B'}{2\eta} \dot u_j \left[ ( u_k u_l +\frac{1}{2}g_{kl} -\frac{\dot u_k \dot u_l}{\eta} )A' + g_{kl} \frac{R'(r)}{12} \right ].
\end{align*}
The subtraction with $j,k$ exchanged gives 
\begin{align*}
C_{jkl} =\frac{B'}{2\eta}\left[( \dot u_j u_k u_l - \dot u_k u_j u_l)(A'+ \frac{A}{r}) + (\dot u_j g_{kl} - \dot u_k g_{jl})( \frac{A'}{2}+\frac{R'}{12} + \frac{A}{r})\right ].
\end{align*}
With the identity $ R' = -8\frac{A}{r} -2A' $ the Cotton tensor gains the useful form \eqref{Cotton}.
\end{proof}
\end{prop}
\begin{prop} {\rm \underline{The Bach tensor} (Bach, 1921, \cite{Bach21})} \\ 
With the Weyl tensor $C_{jkl}{}^m$, the Bach tensor is the only algebraically independent one that is invariant for a conformal transformation $g'_{ij}(x) = e^{2\phi (x)}g_{ij}(x)$ in $n=4$ \cite{Szekeres68}:
$$ \mathscr B_{kl} = 2\nabla^j\nabla^m C_{jklm} + R^{jm} C_{jklm} = -\nabla^j C_{jkl} +  R^{jm} C_{jklm} $$ 
where $C_{jkl}$ is the Cotton tensor. It is symmetric, traceless and divergence-free. In the metric \eqref{metric} we find:
\begin{align}
&\mathscr B_{kl} = B_1 u_k u_l + \frac{1}{4} (B_1-B_2) g_{kl} + B_2 \frac{\dot u_k \dot u_l}{\eta} \label{BACH}\\
& B_1+B_2 = - \frac{2B}{r}\left ( A' +\frac{A}{r}\right ) - \frac{2B}{3}\left ( A' +\frac{A}{r}\right )^\prime \label{BACH1}\\
&B_1 - B_2= -\frac{8}{3} AE - \frac{4}{3}\left ( A' +\frac{A}{r}\right )\left ( B' +\frac{B}{r}\right ) -\frac{4B}{3}\left ( A' +\frac{A}{r}\right )^\prime \label{BACH2}
\end{align}
\begin{proof} With eq.\eqref{WEYL} and using $E_{kl}u^l=0$, it is:
$$ R^{jm} C_{jklm} 
=A(r) \left [ - (g_{kl}+2 u_k u_l )E_{jm}   \frac{\dot u^j \dot u^m}{\eta}
+ \dot u_l E_{km}\dot u^m + \dot u_k E_{jl} \dot u^j\right ] $$
Now use $E_{jk}\dot u^k = \frac{2}{3}E(r) \dot u_l$. Then:
\begin{align}
R^{jm} C_{jklm} =- \frac{4}{3}A(r) E(r) \left [ u_k u_l + \frac{1}{2} g_{kl} - \frac{\dot u_k \dot u_l}{\eta}   \right ]  \label{RICCIWEYL}
\end{align}
\noindent
The divergence of the Cotton tensor is
\begin{align}
\nabla^j C_{jkl} =& \frac{B'}{2\eta}\frac{d}{dr} \left[\frac{B'}{2\eta} \left ( A' +\frac{A}{r}\right ) \right ] \dot u^j \left [ \dot u_j (u_k u_l + \frac{1}{3} g_{kl} ) - (u_j u_l +\frac{1}{3} g_{jl} )\dot u_k  \right ] \nonumber\\
&+ \frac{B'}{2\eta} \left ( A' +\frac{A}{r}\right ) \nabla^j \left [ \dot u_j (u_k u_l + \frac{1}{3} g_{kl} ) -  (u_j u_l +\frac{1}{3} g_{jl} )\dot u_k  \right ] \nonumber\\
=& \frac{1}{3}\left ( A' +\frac{A}{r}\right ) \left[ 2B' (u_k u_l +\frac{1}{2} g_{kl} -\frac{\dot u_k \dot u_l}{\eta}) +\frac{B}{r} 
(5u_k u_l + g_{kl} +\frac{\dot u_k \dot u_l}{\eta})\right ]\nonumber \\
&+ B\left ( A' +\frac{A}{r}\right )^\prime \left [u_k u_l + \frac{1}{3}g_{kl} - \frac{1}{3} \frac{\dot u_k \dot u_l}{\eta}\right ].\label{NABLACOTTON}
\end{align}
It turns out that $\mathscr B_{kl} $ is a traceless linear combination of $u_k u_l$, $g_{kl}$ and $\dot u_k \dot u_l$. The expressions $B_1\pm B_2$
result from the scalars $\mathscr B_{jk}u^j u^k$ and $\mathscr B_{jk}\dot u^j \dot u^k$ evaluated with \eqref{RICCIWEYL} and \eqref{NABLACOTTON}.
\end{proof}
\end{prop}

Now we pin down some space-times that solve special geometric constraints. %
\begin{prop}\label{CASES}
A spherically symmetric static space-time\\
a) has \underline{\sf zero scalar curvature} $(R=0)$ if $2\frac{1-B}{r^2} - 4\frac{B'}{r} - B'' =0$ i.e.
\begin{align}
 B(r) = \frac{b_{-2}}{r^2}+ \frac{b_{-1}}{r} +1  
 \end{align}
The Ricci tensor \eqref{RicciT} is traceless, with $A(r) = 2 b_{-2}/r^4 $.\\
b) is \underline{\sf conformally flat} $(C_{jklm}=0)$ if $E(r)=0$ i.e. 
\begin{align}
B(r) =1+b_1r + b_2 r^2
\end{align}
c) is \underline{\sf harmonic} $(\nabla^m C_{jklm}=0)$ if
\begin{align}
 B(r) = \frac{b_{-1}}{r} +1 + b_1 r +b_2 r^2 \label{HARMONIC}
 \end{align}
Proof: eq.\eqref{Cotton} gives $A'+A/r=0$ i.e. $A=-b_1/r$. Then:
$ B'' + 2\frac{1-B}{r^2} + \frac{2b_1}{r} =0 $, with the above solution. \hfill $\square $\\
d) is \underline{\sf bi-harmonic} $(\nabla^j C_{jkl}=0)$ if it is harmonic \eqref{HARMONIC}, or if
\begin{align}
B(r) = \kappa r^2 \label{BIH}
\end{align} 
with arbitrary constant (note that it s not a special case of harmonic).\\
Proof: The expression \eqref{NABLACOTTON} for $\nabla^j C_{jkl}$ is zero if the components 
$g_{kl}$ and $\dot u_k \dot u_l$ vanish (the component $u_k u_l$ vanishes because of the trace
condition). The difference gives:
\begin{align*}
\left (A'+\frac{A}{r} \right ) \left[ -\frac{B'}{3} + \frac{2}{3} \frac{B}{r}\right ] =0
\end{align*} 
The first factor vanishes for harmonic space-times, the other gives $B=\kappa r^2$, that 
also solves the other constraint and sets $A(r)=1/r^2$, $E(r)=1/(2r^2)$. \hfill $\square $\\
e) is \underline{\sf Einstein} if $A(r)=0$ i.e. 
\begin{align}
 B(r) = \frac{b_{-1}}{r} +1 + b_2 r^2  \label{EINSTEINB}
 \end{align}
f) is \underline{\sf Constant Curvature} if $A(r)=0$ and $C_{jklm}=0$. 
\begin{align}
B(r) = 1+ b_2 r^2
\end{align}
The Riemann tensor has the form $R_{jklm} = \frac{R}{12}(g_{jl} g_{km} - g_{jm} g_{kl})$ with
$R= -12 b_2$.
g) has \underline{\sf zero Bach tensor} $(\mathscr B_{kl}=0)$ if 
\begin{align}
B(r) = -\frac{\beta (2-3b_1 \beta )}{r} + (1-3b_1\beta ) + b_1 r + b_2 r^2 \label{MK}
\end{align}
Proof: the Bach tensor \eqref{BACH} is zero if $B_1\pm B_2=0$. 
With $B\neq 0$ the sum gives $ 3\left ( A' +\frac{A}{r}\right )+ r \left ( A' +\frac{A}{r}\right )^\prime =0 $ i.e  $[r^2 (Ar)^\prime]^\prime =0$,
$A(r) = \frac{a_{-2}}{r^2} +  \frac{a_{-1}}{r}$. Then 
$  B(r) = \frac{b_{-1}}{r} + b_0 + b_1 r + b_2 r^2 $
with $1-b_0=a_{-2}$ and $b_1 = -a_{-1} $. \\
This expression in eq.\eqref{BACH2} gives: $a_{-2}(1+b_0) = 3 a_{-1}b_{-1} $ and $a_{-2}b_1 = - a_{-1}(1-b_0)$.
The first one is the constraint $1-b_0^2=-3b_1 b_{-1}$, the other equation is trivial. \\
A possible parameterization of $B(r)$  is \eqref{MK}. \hfill $\square $
\end{prop}

\begin{prop}[The Faraday tensor]
In a static spherical-symmetric space-time the magnetic coefficient $\mathbb B (r)$ of the Faraday tensor \eqref{FTENSOR} is
\begin{align}
\mathbb B(r) = \frac{q_m}{r^2} \label{MONOPOLE}
\end{align}
where  $q_m$ is a magnetic charge. The current is time-like and independent of $\mathbb B$: 
 \begin{align}
 J^k = (-\eta \varkappa +\mathbb E\sqrt \eta -\frac{\mathbb E}{\sqrt\eta}\nabla_j \dot u^j) u^k 
 \end{align}
\begin{proof}
The equation \eqref{FARADAYCONDQ} for  $\mathbb E/\sqrt\eta $ is satisfied by any function of $r$. The equation \eqref{FARADAYCONDP} for $\mathbb B $ 
becomes
\begin{align*}
\frac{1}{2} \mathbb B' = - \mathbb B \frac{1}{r}.
\end{align*}
with solution \eqref{MONOPOLE}. The expression of the current \eqref{CURRENT} simplifies as directional derivatives other than $\dot u^k\nabla_k$ are zero for scalars that only depend on $r$.
\end{proof}
\end{prop}

The geometric cases presented in Prop.\,\ref{CASES} correspond to well known static spherically symmetric solutions of gravitational theories.\\
We consider the Einstein, the Cotton, the $f(R)$ and the Conformal Gravity theories.

\section{Static solutions in Einstein gravity}
\begin{center}
{\em The imperfect fluid cannot be perfect. Early solutions. Properties of LE and NLE}
\end{center}
The Einstein tensor $G_{jk} = R_{jk}- \frac{R}{2} g_{jk} $ for the static spherical metric \eqref{metric} is
\begin{align}
G_{jk} = \frac{2}{3} A(r) u_j u_k +  \left [\frac{B''}{3} + \frac{B'}{r} 
-\frac{1-B}{3r^2}\right ] g_{jk} 
 - A(r) \left[ \frac{\dot u_j\dot u_k}{\eta} -\frac{u_ju_k+g_{jk}}{3}\right ]. \label{GTENSOR}
 \end{align}
Its tensor structure and the Einstein equation $G_{jk}=T_{jk}$ dictate that of the energy-momentum density $T_{jk}$. 
In the picture of a fluid it is:
\begin{align}
 T_{jk} = (p+\mu) u_j u_k + p g_{jk} + (p_r-p_\perp)  \left[ \frac{\dot u_j\dot u_k}{\eta} -\frac{u_ju_k+g_{jk}}{3}\right ]  \label{TENSOR}
 \end{align}
The structure of the stress tensor is fully specified.
The energy density $\mu$, the effective pressure $p=\frac{1}{3}(p_r+2p_\perp)$, the radial and transverse pressures $p_r$, $p_\perp$ are 
functions $r$:
\begin{align}
\mu = -p_r = \frac{1-B}{r^2} - \frac{B'}{r}, \quad p_\perp = \frac{B''}{2} + \frac{B'}{r} 
\end{align}

Pressure isotropy ($p_r=p_\perp$) imposes $A(r)=\frac{B''}{2} + \frac{1-B}{r^2}=0$, i.e. the space-time is Einstein 
with fluid equation of state $\mu = - p$, and
\begin{align}
 B(r) = 1+\frac{b_{-1}}{r} + b_2 r^2  \label{BBB}
 \end{align} 
The field equation with a dust source, $G_{jk} = \mu u_j u_k$ does not admit a static solution \eqref{metric} (the source 
term must contain a pressure anisotropy to compensate the equality).

\begin{remark}
There has been a discussion whether a static spacetime with spherical metric \eqref{metric} may host a perfect fluid.  Faraoni \cite{Faraoni} and 
Visser \cite{Visser20} showed the inconsistency of the Kiselev metric with a perfect fluid source. A definite negative answer has been given by Lake and Bisson
\cite{Lake19}\cite{Bisson23}. Here, again, we have shown that it does not occur unless $p=-\mu$.
\end{remark}

For the static anisotropic fluid tensor \eqref{TENSOR}, the equation $\nabla_j T^j{}_k =0$ is:
\begin{align*}
0&= (p+\mu) \dot u_k +\nabla_k p  +  (p_r-p_\perp)  \left[\nabla_j \frac{\dot u_j\dot u_k}{\eta} -\frac{\dot u_k}{3}\right ] 
+  \left[ \frac{\dot u_j\dot u_k}{\eta} -\frac{g_{jk}}{3}\right ] \nabla_j (p_r-p_\perp)\\
&=(\mu +p_\perp) \dot u_k +\nabla_k p_\perp +  (p_r-p_\perp)\nabla_j \frac{\dot u^j\dot u_k}{\eta} 
+  \frac{\dot u_k}{\eta}  \dot u^j \nabla_j (p_r-p_\perp)
\end{align*}
A gradient is evaluated in \eqref{NABLAXXX}: $ \nabla_j \frac{\dot u^j \dot u_k}{\eta} =  \dot u_k + \frac{B'}{r\eta} \dot u_k $. 
In spherical symmetry: $\nabla_k p_\perp = \frac{1}{\eta}\dot u_k \dot u^j \nabla_j p_\perp $. The derivative
of the radial pressure is obtained:
\begin{align}
0=p'_r+ \frac{B'}{2B}(\mu +p_r )  + \frac{2}{r}  (p_r-p_\perp)  \label{DIVT}
\end{align}

\subsection{Some static solutions of the Einstein equations} Simple conditions provide the classical static solutions

\noindent
$\bullet$ \underline{\em Schwarzschild} space-time.\\ 
$R_{jk}=0$ gives the famous vacuum spherical solution \eqref{BBB}, with $b_2=0$: 
$$ B(r) = 1 - \frac{2M}{r} $$
\noindent
$\bullet$ \underline{\em Schwarzschild - de Sitter} (SdS) space-time. \\
The equation $G_{jk} +\Lambda g_{jk}=0$ is solved by \eqref{BBB} 
with free parameter $b_{-1}$. The coefficient of $g_{jk}$ fixes $b_2=-\frac{1}{3}\Lambda $.\\
\noindent
$\bullet$ \underline{\em Reissner-Nordstr{\o}m} space-time \cite{DeFelice}.\\  
If $R=0$ the Einstein equation $R_{kl}=T_{kl}$ implies the energy-momentum 
tensor $T_{jk} = \frac{b_{-2}}{r^4}\left [ u_j u_k -\frac{1}{2}g_{jk} -\frac{\dot u_j \dot u_k}{\eta}\right ]$. 
In comoving coordinates the non-zero Faraday component 
$$F_{tr} = \frac{\mathbb E}{\sqrt\eta}u_0 \dot u_r = \mathbb E (-\sqrt B) \frac{1}{2} \frac{B'}{B} = -\frac{\sqrt{b_{-2}}}{r^2} $$
corresponds to the radial electric field of a point charge $q_e=\sqrt{b_{-2}}$.\\
The metric function is (Reissner 1916, Nordstr\"om 1913):
\begin{align}
 B(r) = 1 - \frac{2M}{r} + \frac{q_e^2}{r^2} 
 \end{align} 
%
\noindent
$\bullet$ \underline{\em Reissner-Nordstr{\o}m-(anti) de Sitter}  space-time \cite{Lake79}. \\
It is a variant of the previous metric where 
a cosmological term $-\Lambda g_{jk}$ is added to the traceless $T_{jk}^{em}$. The scale function is:
 $$ B(r) = 1 - \frac{2M}{r} + \frac{q_e^2}{r^2} - \frac{1}{3} \Lambda r^2 $$ 
 
\subsection{Linear and non-linear electrodynamics in Einstein gravity}
While $\mathbb B(r)$ is fixed and equal to $q_m/r^2$ by the Faraday conditon in spherical symmetry, $\mathbb E(r)$ is model dependent.
In linear $(\mathscr L_F=1)$ or non-linear electrodynamics:

\begin{prop} $\mathbb E(r)$ solves the implicit equation (see \cite{Halilsoy15} and \cite{Bronnikov17})
 \begin{align}
\mathbb E (r) = \frac{q_e}{r^2 \mathscr L_F (F)} \label{EQ4Q}
\end{align}
\begin{proof}
In spherical symmetry the equation \eqref{NLE1} for $\mathbb E$  is 
$$\frac{B'}{2} \frac{d}{dr}\log \left [\tfrac{\mathbb E}{\sqrt\eta}\mathscr L_F \right ] =\frac{B'^2}{4B} - \frac{B''}{2} - \frac{B'}{r}$$ 
Then $ \frac{d}{dr}\log [\frac{\mathbb E}{\sqrt\eta}\mathscr L_F]  = \frac{d}{dr}\log \frac{\sqrt B}{B' r^2}$. The integration yields
a constant $q_e$.
\end{proof}
\end{prop}
\noindent
In linear electrodynamics eq.\eqref{EQ4Q} is solved by a Coulomb field
\begin{align}
\mathbb E^{lin}(r) = \frac{q_e}{r^2}
\end{align}
and the electromagnetic energy-momentum density tensor is
\begin{align}
T_{jk}^{lin} = 2\, \frac{q_e^2+q_m^2}{r^4}  \left [ u_j u_k + \frac{1}{2} g_{jk} -\frac{\dot u_j \dot u_k}{\eta} \right ].  \label{TLIN}
\end{align}
This result recovers a generalization of Birkhoff's theorem, stating that a spherical symmetric solution of the Einstein-Maxwell equations is necessarily a piece of the Reissner-Nordstr{\o}m geometry with monopole charges
(see \cite{Gravitation}, p. 844).

The expression $T_{jk}^{nlin}$ in the Einstein equation \eqref{EINSTEINEQ} with the Einstein tensor \eqref{GTENSOR} gives 
eq.\eqref{FFF2} and 
\begin{align}
2 (\mathbb B^2+\mathbb E^2) {\mathscr L}_F (F) = A(r) \label{ATOELLE}
\end{align}

The case $R=0$ i.e. ${\mathscr L}_F=1 $ (linear electrodynamics) is the Reissner-N\"ordstrom
solution with "dyonic charge", i.e. $b_{-2} = q_m^2 + q_e^2$. 

Eq.\eqref{ATOELLE} has been exploited to infer the Lagrangian $\mathscr L$ from the metric function $B(r)$ (through $A(r)$), or the opposite, in two situations: $\mathbb E=0$ or $\mathbb B=0$. The feasibility of the correspondence has been investigated 
by Bronnikov \cite{Bronnikov18}. \\
$\bullet $ Purely magnetic ($\mathbb E=0$). Then $4F  \mathscr L_F (F)=A(r)$ with $F=q_m^2/(2r^4)$. \\
E.~Ay\'on-Beato and A.~Garcia \cite{Ayon00} started with the metric of the Bardeen black-hole, and deduced the Lagrangian:
$$ B(r) =1- \frac{2mr^2}{(r^2 + q_m^2)^{3/2}}\quad \to\quad  
\mathscr L(F) = \frac{3m}{q_m^3} \left [ \frac{\sqrt{2q^2_mF}}{1+ \sqrt{2q^2_mF}} \right ]^{5/2}$$
S.~Kruglov \cite{Kruglov21} obtained the Lagrangian of the Hayward black-hole \cite{Hayward96}:
$$ B(r) =1- \frac{2mr^2}{r^3 + q_m^3}\quad \to\quad  
\mathscr L(F) = \frac{3}{2^{3/4}} \frac{(2q_m^2F)^{3/2}}{(1+ (2q^2_mF)^{3/4})^2 }$$
\noindent
$\bullet $ Purely electric ($q_m=0$). Then $A(r) = - 4F  \mathscr L_F (F)$ with $F=-\mathbb E^2/2$. \\
With the aid of eq.\eqref{EQ4Q} Halilsoy et al. \cite{Halilsoy15} obtained the metric from the Lagrangian:
$$ \mathscr L (F) = \frac{a}{b\sqrt 2 - \sqrt{-4F}}\quad  \to \quad  A(r) = \frac{2bq_e}{r^2} - \sqrt{\frac{|a|q_e}{\sqrt 2}} \frac{2q_e}{r} $$
The function $A(r)$ has the same form as the Mannheim-Kazanas solution \eqref{MK} Prop.\ref{CASES}. 
The same metric function $B(r)$ is also a vacuum solution of Conformal Gravity.\\
An interesting application has been the forecast of the shadow of the black hole in M87 \cite{Kruglov20}.

\section{Linear and non-linear electrodynamics in f(R) gravity}
\begin{center}
{\em The equations and the charged solution by Hollenstein and Lobo}
\end{center}
$f(R)$ gravity is an extension of Einstein gravity, where a function $f(R)$ replaces $R$ in the Einstein-Hilbert action.
The equations in spherical symmetry are studied by Capozziello et al. in \cite{Capozziello08}.
With coupling to non-linear electrodynamics, the equations of motion, with $f_R=df/dR$ are \cite{Hollenstein08}:
\begin{align}
& R_{jk} f_R(R) -\frac{1}{2}g_{jk}f(R) + [g_{jk} \square  -\nabla_j\nabla_k]  f_R(R) = T^{nlin}_{jk}\\
&\nabla_j  (F^{jk}  \mathscr L_F(F)) =0
\end{align}
The second one is the same as in the Einstein theory. The equations are studied in the static metric \eqref{metric}.
For any function $g(r)$: $\nabla_k g=  \dot u_k \frac{2B}{B'} g' $ (eq. \eqref{GRAD}), and
\begin{align*}
\nabla_j \nabla_k g =& (\nabla_j \dot u_k) (\frac{2B}{B'} g') + \dot u_j \dot u_k \frac{2B}{B'}\frac{d}{dr}(\frac{2B}{B'} g')\\
 =& u_j u_k \left[ \frac{B}{r} -\frac{B'}{2}\right ]g' + g_{jk} \frac{B}{r}g' + \frac{\dot u_j \dot u_k}{\eta} 
\left[ Bg'' +\frac{1}{2}B' g' - \frac{Bg'}{r}\right ]
\end{align*}
where $\nabla_j \dot u_k$ is \eqref{nabladot}. In particular: $\square g = 2\frac{B}{r}g' +B'g'+ Bg'' $.\\
Given the expressions of the Ricci tensor \eqref{RicciT}  and of $T_{jk}^{nlin}$
\eqref{NONLINENMOMTENS}, the first field equation corresponds to three scalar equations: 
\begin{align}
&-\frac{1}{2}f + f_R \left[ \frac{R}{4}+\frac{1}{2}A(r)\right ] +\square f_R -  \frac{B}{r} f_{R}' = 2[
\mathbb B^2 \mathscr L_F -\mathscr L] \label{fR1}\\
&f_R A(r) +\left [\frac{B'}{2} -\frac{B}{r}\right ] f_{R}' = 2(\mathbb E^2+\mathbb B^2) \mathscr L_F\label{fR2}\\
&f_R A(r) + \left[ \frac{B'}{2} - \frac{B}{r}\right ] f_{R}'  + Bf_R''  = 2(\mathbb E^2+\mathbb B^2) \mathscr L_F \label{fR3}
\end{align}
The difference of equations \eqref{fR2} and \eqref{fR3} is $f_R'' =0$. Thus, 
we reobtain a simple general result by Hollenstein and Lobo \cite{Hollenstein08}:
\begin{prop} 
In $f(R)$-nonlinear electrodynamics with static metric \eqref{metric}, it is $f_R(R(r)) = cr + d  $,
where $c$ and $d$ are constants.
\end{prop}
\noindent
The case $c=0$, $d=1$ is Einstein's gravity ($f=R$). \\
The result greatly simplifies equations \eqref{fR1} and \eqref{fR2}. With $\frac{1}{4}R+\frac{1}{2}A=
\frac{1-B}{r^2} -\frac{B'}{r}$ and  $\square f_R = 2\frac{B}{r}c +B'c$, they become:
\begin{align}
&\frac{1}{2}f(R) = (cr+d) \left [ \frac{1-B}{r^2} -\frac{B'}{r}\right ] + c\left [ \frac{B}{r} + B'\right ] -  2[
\mathbb B^2 \mathscr L_F -\mathscr L]  \label{EQQQ1}\\
&(cr+d) \left[ \frac{1-B}{r^2} + \frac{B''}{2}\right ] + c \left[ \frac{B'}{2}- \frac{B}{r}\right] = 2(\mathbb E^2+\mathbb B^2) \mathscr L_F
\label{EQQQ2}
\end{align} 
The spherical symmetry always forces $\mathbb B=q_m/r^2$. To go further, we consider linear electrodynamics 
$\mathscr L_F=1$, $\mathbb E=q_e/r^2$. Eq.\eqref{EQQQ2} can now be solved and is
eq.34 in \cite{Hollenstein08} (since it is without explanation, we offer a derivation in Appendix 3).
\begin{align}
B(r) = 1& +\frac{cK}{2d^2} -\frac{K}{3dr}- \left[1+\frac{cK}{d^2} + 4 \frac{q_e^2+q_m^2}{d} (\frac{c}{d})^2\right] \left[\frac{c}{d} r - (\frac{c}{d})^2 r^2 \log \frac{cr+d}{r} \right ] \nonumber \\
&+  \frac{q_e^2+q_m^2}{d}\left [ \frac{1}{r^2} - (\frac{c}{d}) \frac{4}{3r} + 2(\frac{c}{d})^2  \right ] + K_0r^2. \label{LOOB}
 \end{align}
The solution $B(r)$  has to produce in \eqref{EQQQ1} a function $f(R)$, and be compatible with $f_R(R)=cr+d$.\\
The thermodynamics of a $f(R)=R-2\alpha \sqrt R $ black hole with metric \eqref{metric} are studied in \cite{Nashed19}, in power law electrodynamics. 

%
%
%
\section{Static solutions in Cotton gravity}
\begin{center}
{\em Cotton gravity is Einstein gravity with a free Codazzi tensor.\\ Two new solutions: perfect fluid and LE.}
\end{center}
Cotton gravity was introduced by Harada \cite{Harada}, as an extension of Einstein's gravity. 
In the Harada equation, the Einstein tensor is replaced by the Cotton tensor, and the energy-momentum tensor is replaced by gradients of it:
\begin{align}
C_{jkl} = \nabla_j T_{kl} - \nabla_k T_{jl} - \frac{1}{3}(g_{kl}\nabla_j T - g_{jl}\nabla_k T) \label{COTTONFIELDEQ}
\end{align}
where $T=T^k{}_k$. As we showed in \cite{Codazzi} the Harada equation is equivalent 
to the Einstein equation with an energy momentum modified by an arbitrary Codazzi tensor
\begin{align}
& R_{jk} - \tfrac{1}{2}g_{jk} R  = T_{jk} +\mathscr C_{jk} - g_{jk} \mathscr C^k{}_k\\
& \nabla_i \mathscr C_{jk} = \nabla_j \mathscr C_{ik}\nonumber
\end{align}
\subsection{The Harada solution} Harada found a static spherical solution of his equation with $C_{jkl}=0$. It is \eqref{HARMONIC} in Prop.\,\ref{CASES}:
\begin{align}
 B(r) = 1 - \frac{2M}{r} - \frac{\Lambda}{3}r^2  +\gamma r  \label{HARADA}
 \end{align}
 It generalizes the Schwarzschild solution by a cosmological term, and corresponds to solving
 the Einstein equation with the energy momentum
$$ T_{jk} =-\Lambda g_{jk}+ \frac{\gamma}{r} \left [- \frac{2}{3} u_j u_k  + \frac{4}{3}  g_{jk}
 + \left ( \frac{\dot u_j\dot u_k}{\eta} -\frac{u_ju_k+g_{jk}}{3}\right )\right ] $$
 Therefore, the Harada vacuum solution is a solution of the Einstein equation for an exotic anisotropic fluid, 
 with velocity $u_k$, energy density $\mu =-p_r= -\frac{2\gamma}{r} +\Lambda $ and transverse pressure
 $p_\perp = \frac{\gamma}{r} - \Lambda$.\\
 The same function $B(r)$ appears as solution of a model for gravity at large distances studied by Grumiller \cite{Grumiller}, with an analogous energy-momentum tensor.
 
Harada numerically solved the equations for Cotton-gravity to describe the rotation curves of galaxies \cite{Harada22}, where a linear term $\gamma r$ provides the observed gravitational 
 potential without the need of dark matter.

\subsection{Perfect fluid solution}
While in Einstein gravity there are no perfect fluid solutions with the static spherical metric \eqref{metric}, this is no longer true in Cotton gravity because of the freedom of choosing the Codazzi tensor. \\
The following one, with constants $K$ and $\kappa$, 
\begin{align}
\mathscr C_{jk} =  \tfrac{K}{\sqrt {B(r)}} u_j u_k +\kappa g_{jk} \label{CodazziPerfect}
\end{align}
is a Codazzi tensor in the metric \eqref{metric} (see \cite{Codazzi}). By choosing
$B(r)= \frac{b_{-1}}{r} +1 + b_2 r^2 $ (as \eqref{EINSTEINB}, i.e. $A(r)=0$), the Ricci tensor is Einstein, $R_{jk}= - 3 b_2 g_{jk}$. The energy-momentum tensor 
\begin{align*}
T_{jk} &= R_{jk} - \tfrac{1}{2}R g_{jk} - \mathscr C_{jk} + g_{jk} \mathscr C^k{}_k \\
&=-\tfrac{K}{\sqrt{B(r)}} (u_j u_k +g_{jk}) - \left[\tfrac{R}{4}+ 4\kappa\right ] g_{jk} 
\end{align*}
is perfect fluid, and the Harada equation \eqref{COTTONFIELDEQ} is solved by the metric \eqref{metric} with
the function $B(r)$.  The perfect fluid
has constant energy density $\mu=4\kappa  +R/4$, while $p+\mu=-\frac{K}{\sqrt B(r)}$ is a function of $r$ because of the Codazzi term.\\
- if $\mathscr C_{jk}=0$ we recover GR with the cosmological law $p=-\mu$.\\
- if $K =0$, $\kappa = -\frac{R}{16}$ we get the empty solution $p=\mu=0$. Thus in Cotton gravity the same metric (SdS or SadS) is compatible with different  energy-momentum tensors. \\
- The electric function is $E(r)=-\frac{3}{2}\frac{b_{-1}}{r^3}$. If $b_{-1}=0$ then $C_{jklm}=0$. 
$B(r)= 1+b_2 r^2$ gives the metric of a constant curvature space-time
$R_{jklm}=\frac{1}{12}R(g_{jl}g_{km}- g_{jm}g_{kl})$ in presence of a perfect fluid in Cotton gravity.
\begin{remark}
Apparently, the statement that \eqref{CodazziPerfect} is a Codazzi tensor in a constant curvature space-time comes at odds with the theorem by Ferus \cite{Ferus} stating that the only Codazzi tensors is such spacetimes are $\nabla_j\nabla_k \varphi + \frac{1}{12}R \varphi g_{jk}$, where $\varphi $ is an arbitrary scalar field. Actually, it can be shown that the tensor is in this class with $\varphi (r) = K \sqrt{B(r)}$. The term $\kappa g_{jk}$ is the trivial Codazzi tensor.
\end{remark}
 
\subsection{Linear electrodynamics}
We obtain a new solution for Cotton gravity in presence of the linear tensor of electrodynamics.
\begin{prop}  The metric function $B(r)$ solving the Cotton gravity equation in linear electrodynamics is:
\begin{align}
B(r) = \left[ 1- \frac{2M}{r} -\frac{\Lambda}{3}r^2 +\gamma r \right ]+ \frac{q_e^2 + q_m^2}{r^2}  
\end{align}
It is the sum of the solution of $C_{jkl}=0$ (in square brackets) and a dyonic charge term.
\begin{proof}
The traceless energy-momentum tensor $T_{jk}^{lin}$ in eq.\eqref{LINENMOMTENS}, is entered in the Cotton gravity  equation: $C_{jkl}= \nabla_j T_{kl}^{lin} - \nabla_k T_{jl}^{lin}$:
\begin{align}
C_{jkl} = \left [K'+\frac{K}{r}\right] (u_k \dot u_j - u_j \dot u_k) + \left[ \frac{K'}{2} + \frac{K}{r}\right ] (\dot u_j g_{kl}- 
\dot u_k g_{jl})
\end{align}
where for brevity $K=2[\frac{q_m^2}{r^4} +\mathbb E^2(r)] $. The static spherical Cotton tensor is \eqref{Cotton}. \\
The contraction with $g^{kl}$ gives $0=\frac{K'}{2}+\frac{2K}{r}$, with solution 
$$\mathbb E(r)=\frac{q_e}{r^2} $$ 
The contraction with $u^k u^l$ is $ A' + \frac{A}{r} = \frac{3}{4} K'$ with solution
$A(r) = -\frac{\gamma}{r} + 2\frac{q_e^2+q_m^2}{r^4}$, with a constant $\gamma$. The corresponding metric function is obtained.
\end{proof}
\end{prop}

\section{Static solutions in Conformal gravity}
\begin{center}
{\em The field equations, the Mannheim-Kazanas and LE solutions.}
\end{center}
The action of conformal gravity is
$S =  -\alpha_G \int d^4x \sqrt{(-g)} C_{jklm} C^{jklm} + S_{matter}$
In $n=4$ the Weyl term, that accounts for geometry, is invariant for the conformal transformation\footnote{It is $C'_{jklm} =e^{2\phi} C_{jklm}$,
$C'^{jklm} = e^{-6\phi}C_{jklm} $ and $\sqrt{-g'}= e^{4\phi} \sqrt {-g}$.} $g'_{jk}(x) = e^{2\phi (x)} g_{jk}(x)$. \\
The variation in the metric tensor, neglecting boundary terms, is:
$$ \delta S= 2\alpha_G \int d^4x \sqrt{(-g)} \mathscr B_{kl}\, \delta g^{kl} - \frac{1}{2}\int d^4x \sqrt{(-g)}  \,
T_{kl}\, \delta g^{kl} $$
where $\mathscr B_{kl} = 2\nabla^j\nabla^m C_{jklm} + R^{jm} C_{jklm} = -\nabla^j C_{jkl} + R^{jm} C_{jklm} $ is the Bach tensor and $T_{kl}$ is the energy-momentum density tensor. 
The field equation of Conformal gravity is: 
\begin{align}
4\alpha_G \, \mathscr B_{kl} = T_{kl} \label{CFE}
\end{align} 
The property $\nabla_j T^j{}_k=0$ is mantained by the identity  $\nabla_j \mathscr B^j{}_k=0$.

Eq.\eqref{BACH} for the static spherical Bach tensor fixes the form of the energy-momentum tensor as an anisotropic fluid \eqref{TENSOR}:
$$ 4\alpha_G \left [B_1 u_j u_k +\frac{B_1-B_2}{4} g_{jk} +B_2 \frac{\dot u_j\dot u_k}{\eta} \right ]= 
 (\mu +p_\perp) u_j u_k + p_\perp g_{jk} + (p_r-p_\perp)   \frac{\dot u_j\dot u_k}{\eta}  $$
 with 
$\mu = \alpha_G (3B_1 + B_2)$, $p_r = \alpha_G(B_1+3B_2)$ and $p_\perp = \alpha_G (B_1-B_2)$.\\
Since the Bach tensor is traceless, it is $T^k{}_k=0$, i.e. in static conformal gravity the fluid always satisfies
\begin{align}
 p_r+ 2p_\perp  =\mu 
 \end{align}
 The continuity equation for the energy momentum is eq.\eqref{DIVT}.
 
 Let us view some special cases:
\subsection{Vacuum solution} Mannheim and Kazanas \cite{MannKaz} obtained the vacuum spherical static
solution for conformal gravity, $\mathscr B_{jk}=0$. It is the metric function $B(r)$ in eq.\eqref{MK} Prop.\,\ref{CASES}.
The solution arose much interest for the description of the rotation curves of galaxies, where the linear term
$\gamma r$ accounts for the plateau without need of dark matter \cite{MannKaz} \cite{Mannheim12} \cite{Horne16} \cite{Hobson21}. 
Constraints on the value of the constant $\gamma $ were obtained by Sultana et al.
\cite{Sultana12}, using data for perihelion shift.

Bach showed that every static spherically symmetric space-time that is conformally related to the Schwarzschild-de Sitter (SdS) metric solves $\mathscr B_{jk}=0$ \cite{Mann91}. The converse was later proved by Buchdahl (see \cite{Hobson21}). In some papers it is actually proven that \eqref{MK} is conformally equivalent to the SdS metric.

\subsection{Perfect fluid} The anisotropic term is zero if $B_2=0$, and $p_r=p_\perp =\mu/3$.\\
 The condition on $B_2$ is a fourth order non-linear differential equation for $B(r)$.
%
\subsection{The anisotropic EoS $\boldsymbol{\mu = -p_r}$} This occurs for $B_1+B_2=0$. The trace-less energy momentum tensor takes the same form of $T_{jk}^{lin}$ of linear electrodynamics, eq.\eqref{LINENMOMTENS}:
$$ T_{kl} = \mu \left [ u_k u_l + \frac{1}{2} g_{kl} - \frac{\dot u_k \dot u_l}{\eta} \right ] $$ 
For this reason, the case is by far the most studied in the literature.
Remarkably, Riegert \cite{Riegert84} proved that Birkhoff's theorem holds in conformal gravity and implies that a spherical symmetric solution of the Bach-Maxwell equations is necessarily static, with $B(r)$ given below.

Eq.\eqref{BACH1} gives:
$0=  \frac{1}{r}\left ( A' +\frac{A}{r}\right ) + \frac{1}{3}\left ( A' +\frac{A}{r}\right )^\prime $, 
with solution $A(r) =\frac{1}{r^2}(1-b_0) - \frac{1}{r} b_1$. The equation has solution
$$ B(r) = \frac{b_{-1}}{r} + b_0 + b_1 r + b_2 r^2 $$ 
It follows that $E(r)=\frac{1}{2r^2}(1-b_0^2)-\frac{3}{2} \frac{b_{-1}}{r^3}$. This, in eq.\eqref{BACH2} gives 
$2B_1 =  -\frac{4}{3r^4}[(1-b_0^2) +3b_1b_{-1}]$. 
The energy density is $\mu=p_\perp=-p_r= 2\alpha_G B_1$:
$$\mu (r) =  -\alpha_G \frac{8}{3r^4}[(1-b_0^2) +3b_1b_{-1}]$$
The dependence $r^{-4}$ agrees with the monopole field in linear electrodynamics. In this picture: 
$$ 2(q_e^2 + q_m^2)  = -\alpha_G \frac{8}{3}[(1-b_0^2) +3b_1b_{-1}] $$
The parameter $b_2$ is free while $b_{\pm 1}$ and $b_0$ are constrained. The metric function $B$ can be cast as
follows \cite{Mann91}
$$ B(r) = -\frac{1}{r}\left [\beta (2-3\beta\gamma) +\frac{q_e^2+q_m^2}{4\gamma \alpha_G }\right ] + (1-3\beta\gamma) + \gamma r -\kappa r^2 $$
Let's look at two subcases:
\subsubsection*{The harmonic solution ($\nabla_mC_{jkl}{}^m=0$)} The function $B(r)$ by Harada eq.\eqref{HARADA} solves $C_{jkl}=0$. It implies $A(r) =-\gamma/r$ and  $E(r) = 3M/r^3$, i.e. $b_0=1$, $b_1=\gamma $ and $b_{-1}=-2M$. 
Therefore: the Harada metric \eqref{HARADA} solves the field equation of conformal gravity in presence of electric and magnetic monopoles, with charge $q_e^2+q_m^2 = 8 M \gamma \; \alpha_G$.
%
\subsubsection*{The bi-harmonic solution  ($\nabla_j\nabla_mC^j{}_{kl}{}^m=0$)} Besides the harmonic solution, the  function $B(r)=\kappa r^2$, see \eqref{BIH}, cancels the term $\nabla^j C_{jkl}$. With $b_0=0$, $b_1=b_{-1}=0$, 
$B(r)=\kappa r^2$ solves the field equation of conformal gravity coupled to monopole charges $q_e^2+q_m^2 =-\frac{4}{3} \alpha_G$, independent of $\kappa $, with $\alpha_G<0$.\\

Some equations of state $p_r=p_r(\mu)$ have been numerically studied by Brihaye and Verbin \cite{Brihaye09}.

\section{Conclusions}
The initial effort of writing tensors with the vectors $u_j$, $\dot u_j$ that define static space-times, and two other orthogonal vectors, is rewarded by the simplicity of the study of the field equations in gravitation theories. 
In spherical symmetry the first two vectors suffice, the others being projected away with entrance of the metric tensor. 
In the field equations, a geometric tensor equals a matter tensor; the tensor form of the first determines that of the latter, and the equality of the coefficients are scalar field equations.\\
With this plan we obtain a list of solutions in Einstein, Cotton, $f(R)$ and conformal gravity, with results on the Faraday tensor and (non)linear - electrodynamics, and new solutions in Cotton gravity. \\
New and old results are here obtained in the natural and simple covariant formalism. \\
This strategy may be applied to other extended theories, as Gauss-Bonnet gravity.

\section*{Appendix 1}
We report some useful formulas valid for static spherical space-times. They result from the equations for
doubly-warped spherical space-times with $a(t)=1$ presented in ref.\cite{DWST}. In this paper $b^2=B$, $f_1^2=1/B$, $f_2^2=r^2$ and $n=4$.
\begin{align}
ds^2  = -b^2(r) dt^2 + f_1^2(r) dr^2 + f_2^2(r) d\Omega_{n-2}^2 \label{sph}
\end{align}
The Ricci tensor is eq.49 in \cite{DWST}. Here $\xi =0$. It is the sum of a perfect fluid term and a traceless tensor:
\begin{align}
& R_{kl} = \frac{R+n\nabla_p\dot u^p}{n-1} u_k u_l  + \frac{R+\nabla_p\dot u^p}{n-1} g_{kl} + 
\Sigma (r) \left [\frac{\dot u_k \dot u_l}{\eta} - \frac{u_k u_l+g_{kl}}{n-1}\right ]\\
&\Sigma (r) = \nabla_p \dot u^p  
-(n-1)\left( \eta + \frac{\dot u^p\nabla_p\eta}{2\eta}\right ) -(n-2) E(r).\nonumber
\end{align}
While in general $u^l$ is an eigenvector, with spherical symmetry also $\dot u^l$ is an eigenvector.
The electric tensor (eqs. 48 and 44 in \cite{DWST}) is
\begin{align}
&E_{kl} = E(r)  \left [\frac{\dot u_k \dot u_l}{\eta} - \frac{u_k u_l+g_{kl}}{n-1}\right ] \label{ELECTRIC2}\\
&E(r) = \frac{n-3}{n-2}\frac{1}{f_1^2} \left[ \frac{f_1^2}{f_2^2} + \frac{f_2''}{f_2}-\frac{f'_2{}^2}{f_2^2} 
-\frac{f_1'f_2'}{f_1f_2}+\frac{b'f_1'}{bf_1} +\frac{b'f_2'}{b f_2} - \frac{b''}{b}\right ] \nonumber
\end{align}
\begin{align}
&\nabla_p \dot u^p = \frac{1}{bf_1^2}\left[ b'' -b'\frac{d}{dr}\log (f_1f_2) + (n-1)b' \frac{f_2'}{f_2}\right ] \label{AB1}\\
&\frac{\dot u^p\nabla_p \eta}{2\eta}= \frac{1}{f_1} \frac{d}{dr}  \frac{b'}{f_1 b} \label{AB2}
\end{align}
The scalar $\Sigma (r)$ is evaluated with the aid of eq.51 in \cite{DWST}:
\begin{align}
\Sigma (r) =  -\frac{1}{f_1^2} \left [ \frac{b''}{b}-\frac{b'f_1'}{b f_1}-\frac{b' f_2'}{b f_2} \right ] - 
\frac{n-3}{f_1^2} \left[ \frac{f_1^2}{f_2^2} + \frac{f_2''}{f_2}-\frac{f'_2{}^2}{f_2^2}
-\frac{f_1'f_2'}{f_1f_2}\right ].
\end{align}
The curvature scalar of space-time and of the space submanifold are:
\begin{align}
&R = R^\star - 2 \nabla_p \dot u^p \label{SCALARR}\\
&R^\star = \frac{(n-2)(n-3)}{f_2^2} -\frac{n-2}{f_1^2} \left[ 2\frac{f_2''}{f_2} - 2\frac{f_1'f_2'}{f_1f_2} + (n-3) \frac{f_2'^2}{f_2^2}\right ]
\end{align}

\section*{Appendix 2: proof of theorem \ref{THFAR}}
With \eqref{FTENSOR} and Lemma \ref{LEMMAYZ}:
\begin{align}
 \nabla_i F_{jk} 
=&
(\nabla_i \frac{\mathbb E}{\sqrt\eta})(u_j \dot u_k - \dot u_j u_k) + (\nabla_i \mathbb B)(y_j z_k - y_k z_j) 
+\frac{\mathbb E}{\sqrt \eta}(u_j\nabla_i \dot u_k - u_k \nabla_j \dot u_i) \nonumber \\
&+\mathbb B (- Y_i \dot u_j z_k - Z_i y_j \dot u_k  + Y_i z_j\dot u_k + Z_i\dot u_j y_k)  \label{nablaF}
\end{align} 
The cyclic sum is:
\begin{align*}
&(\nabla_i \frac{\mathbb E}{\sqrt \eta})(u_j \dot u_k - \dot u_j u_k)+(\nabla_j \frac{\mathbb E}{\sqrt\eta})(u_k\dot u_i - \dot u_k u_i)+ (\nabla_k \frac{\mathbb E}{\sqrt \eta})(u_i \dot u_j - \dot u_i u_j)\\
&+ (\nabla_i \mathbb B)(y_j z_k - y_k z_j)+(\nabla_j \mathbb B)(y_k z_i - y_i z_k)+(\nabla_k \mathbb B)(y_i z_j - y_j z_i)\\
&+\frac{\mathbb E}{\sqrt \eta}(u_j\nabla_i \dot u_k - u_k \nabla_i \dot u_j +u_k\nabla_j \dot u_ i- u_i \nabla_j \dot u_k +u_i\nabla_k \dot u_j - u_j \nabla_k \dot u_i )\\
&+ \mathbb B Y_i (z_j\dot u_k -z_k \dot u_j) - \mathbb B Z_i (y_j \dot u_k - y_k\dot u_j) +
 \mathbb B Y_j (z_k\dot u_i -z_i\dot u_k) - \mathbb B Z_j (y_k \dot u_i -y_i\dot u_k) \\
&+ \mathbb B Y_k (z_i \dot u_j -z_j\dot u_i) - \mathbb B Z_k (y_i \dot u_j  - y_j\dot u_i) 
\end{align*}
The third line is zero because the acceleration is closed.
For the cyclic sum to be zero, all contractions with vectors must be zero, and give conditions. Contraction with $u^i$ gives:
$(\nabla_j \frac{\mathbb E}{\sqrt\eta}) \dot u_k - (\nabla_k \frac{\mathbb E}{\sqrt \eta}) \dot u_j = 0$ with solution 
$$\nabla_j (\frac{\mathbb E}{\sqrt\eta} )= \varkappa \dot u_j$$
With this result the ciclic condition simplifies: 
\begin{align*}
& (\nabla_i \mathbb B)(y_j z_k - y_k z_j)+(\nabla_j \mathbb B )(y_k z_i - y_i z_k)+(\nabla_k \mathbb B)(y_i z_j - y_j z_i)\\
&+ \mathbb BY_i (z_j\dot u_k -z_k \dot u_j) - \mathbb B Z_i (y_j \dot u_k - y_k\dot u_j) +
 \mathbb B Y_j (z_k\dot u_i -z_i\dot u_k) - \mathbb B Z_j (y_k \dot u_i -y_i\dot u_k) \\
&+ \mathbb B Y_k (z_i \dot u_j -z_j\dot u_i) - \mathbb BZ_k (y_i \dot u_j  - y_j\dot u_i)=0. 
\end{align*}
Contraction with $y^i$:
 \begin{align*}
& (y^i\nabla_i \mathbb B )(y_j z_k - y_k z_j) - (\nabla_j \mathbb B ) z_k+ (\nabla_k \mathbb B ) z_j \\
&+ \mathbb B y^iY_i (z_j\dot u_k -z_k \dot u_j) - \mathbb B y^iZ_i (y_j \dot u_k - y_k\dot u_j) +\mathbb B (Z_j \dot u_k - Z_k \dot u_j )  =0.
\end{align*}
A further contraction with $z^j$ gives:
$\nabla_k \mathbb B  =  (y^i\nabla_i \mathbb B ) y_k  +(z^j\nabla_j \mathbb B ) z_k - \mathbb B (y^jY_j + z^jZ_j )\dot u_k $ i.e.  
$\dot u^k \nabla_k \mathbb B  = - \eta \mathbb B  (y^rY_r +z^r Z_r )$.
The right-hand-side is evaluated in Lemma \ref{LEMMAYZ} and gives the second condition.

Using the form of $\nabla_j \mathbb B $, the cyclic condition becomes 
\begin{align*}
& -(y^rY_r + z^rZ_r ) [\dot u_i  (y_j z_k - y_k z_j)+ \dot u_j (y_k z_i - y_i z_k)+\dot u_k (y_i z_j - y_j z_i) ]\\
&+ Y_i (z_j\dot u_k -z_k \dot u_j) - Z_i (y_j \dot u_k - y_k\dot u_j) +
 Y_j (z_k\dot u_i -z_i\dot u_k) - Z_j (y_k \dot u_i -y_i\dot u_k) \\
&+ Y_k (z_i \dot u_j -z_j\dot u_i) - Z_k (y_i \dot u_j  - y_j\dot u_i) =0.
\end{align*}
The contractions with $\dot u^i$, $y^i$ or $z^i$ or with the metric tensor are trivial. Indeed it is satisfied by the generic 
expansions $Y_i = ay_i + b z_i + c \dot u_i$ and $Z_i = a' y_i + b' z_i + c' \dot u_i$.

$\dot\varkappa = u^k\nabla_k (\frac{1}{\eta}\dot u^j \nabla_j \frac{\mathbb E}{\sqrt\eta})$. Use $\dot \eta=0$, $\ddot u^j =\eta u^j$ and $u^j\nabla_j \frac{\mathbb E}{\sqrt\eta}=0$. Then
$\dot\varkappa =\frac{1}{\eta}  \dot u^j u^k\nabla_k \nabla_j \frac{\mathbb E}{\sqrt\eta} = \frac{1}{\eta}  \dot u^j u^k\nabla_j \nabla_k \frac{\mathbb E}{\sqrt\eta} $. Now: $$u^k\nabla_j \nabla_k \frac{\mathbb E}{\sqrt\eta} =
\nabla_j (u^k\nabla_k \frac{\mathbb E}{\sqrt\eta}) - (\nabla_j u^k)\nabla_k \frac{\mathbb E}{\sqrt\eta} = \nabla_j (\varkappa  u^k \dot u_k) + u_j \dot u^k (\varkappa \dot u_k) = u_j\eta\varkappa $$
Then: $\dot \varkappa = \dot u^j u_j \varkappa =0 $.

\section*{appendix 3: Solution of eq.\eqref{EQQQ2} with point charges.}
After multiplication of \eqref{EQQQ2} by $2r^2$, the equation takes the form $PB'' + QB' + TB = S$ with 
$P(r)= cr^3+dr^2$, $Q(r)=cr^2$, $T(r) = -4cr -2d$. The circumstance $P''-Q'+T=0$ makes the equation integrable (see \cite{Zwillinger} eq.67.5). Indeed it can be written as 
$$[(cr^3+dr^2)B]''-[B(5cr^2 + 4rd)]' = 4 r^2 \frac{q_e^2+q_m^2}{r^4} -2(cr+d)$$
An integration gives a constant $K$: 
$$ B'  - \frac{2}{r} B  = - \frac{cr +2d}{r(cr+d)} + \frac{1}{r^2(cr+d)}\left [ K+4 \int^r dr' \, \frac{q_e^2+q_m^2}{r'^2}\right ]$$
Define $B(r)=r^2 H(r)$. The equation now is:
\begin{align*}
 \frac{dH}{dr}   =& - \frac{cr +2d}{r^3(cr+d)} +\frac{1}{r^4(cr+d)} \left[K - 4 \frac{q_e^2+q_m^2}{r}\right ]\\
 =& -\frac{1}{r^3} + \frac{K}{dr^4} - \frac{d^2+cK}{d}\frac{1}{r^3(cr+d)} - 4 \frac{q_e^2+q_m^2}{r^5 (cr+d)} 
\end{align*}
Note that: $\frac{1}{r^3(cr+d)}= \frac{1}{d} [\frac{1}{r^3} - \frac{c}{d} \frac{1}{r^2} + (\frac{c}{d})^2 \frac{1}{r}] -
(\frac{c}{d})^3 \frac{1}{cr+d}$, and similar with power $5$. The integral gives another constant $K_0$:
\begin{align*}
H(r) =& \frac{1}{2r^2} -\frac{K}{3dr^3}- \frac{d^2+cK}{d^2}\left[- \frac{1}{2r^2} + \frac{c}{d} \frac{1}{r} - (\frac{c}{d})^2 \log \frac{cr+d}{r} \right ] + K_0\\
&-4 \frac{q_e^2+q_m^2}{d}\left [ - \frac{1}{4r^4}
 + (\frac{c}{d}) \frac{1}{3r^3} - (\frac{c}{d})^2 \frac{1}{2r^2} + (\frac{c}{d})^3 \frac{1}{r} - (\frac{c}{d})^4 \log \frac{cr+d}{r}
 \right ] 
\end{align*}
Multiplication by $r^2$ gives the solution $B(r)$ in \eqref{LOOB}. It coincides with eq.34 in \cite{Hollenstein08}. With zero charge it is eq.22 in \cite{Sebastiani11}.

\subsection*{Data availability}
Data sharing is not applicable to this article as no datasets were generated or analyzed during the current study.

\vfill
\end{document}